%% file: main_V2.tex
\documentclass[aps,pra,twocolumn,superscriptaddress]{revtex4-2}

\usepackage{url}
\usepackage[utf8]{inputenc}
\usepackage[T1]{fontenc}

\usepackage{tikz}
\usetikzlibrary{arrows}

\usepackage[normalem]{ulem}
\usepackage{color}
\usepackage{mathtools}
\usepackage{graphicx}
\usepackage{hyperref}
\usepackage{xfrac}

\usepackage{cleveref}
\usepackage{float}
\restylefloat{table}

\usepackage{amsmath,amssymb,amsxtra,amstext,amsfonts,dsfont}
\usepackage{multirow}
\usepackage{makecell}
\usepackage{array}
\usepackage{upgreek}
\usepackage{mathrsfs}
\usepackage{bbold}
\usepackage{bm}

\usepackage{ragged2e}

\usepackage{soul,xcolor}
\setstcolor{red}

\usepackage{physics}
\usepackage{braket}

\usepackage[squaren]{SIunits}

\newcommand{\xzpf}{x_{\text{zpf}}}
\renewcommand{\aa}{\ensuremath{\hat{a}}}
\newcommand{\aad}{\ensuremath{\hat{a}^\dagger}}

\newcommand{\timeedge}{t_\text{r}} 
\newcommand{\timedispeff}{t_{d,\text{eff}}}

\newcommand{\fperiod}{\;.}
\newcommand{\fcomma}{\;,}

\newcommand{\Om}{\Omega}
\newcommand{\displacement}{\alpha}

\renewcommand{\var}[1]{\ensuremath{\text{Var}[#1]}}
\newcommand{\cov}[2]{\ensuremath{\, \text{Cov}[#1,#2] }}
\newcommand{\vect}[1]{{\bm{#1}}}			
\newcommand{\avg}[1]{\left\langle #1 \right\rangle}		

\newcommand{\sqHell}{\mathcal{D}^2_H}

\newcommand*{\hbaraux}[2]{\sbox0{\mathsurround=0pt$#1\mathchar'26$}\mkern-1mu\lower.07\ht0\box0\mkern-8mu}

\makeatletter
\newcommand{\nosubsubsection}[1]{%
  \begingroup\let\addcontentsline\@gobblethree
  \subsubsection*{#1}\endgroup}
\makeatother

\begin{document}

\title{Quantum-enhanced mechanical sensing with machine-learning estimation}

\author{Finn Müßigbrodt}
\affiliation{Department of Physics, ETH Z\"{u}rich, Z\"{u}rich 8093, Switzerland}

\author{Andraž Omahen}
\affiliation{Department of Physics, ETH Z\"{u}rich, Z\"{u}rich 8093, Switzerland}

\author{Q. Rumman Rahman}
\affiliation{Department of Electrical and Computer Engineering, Princeton University, Princeton, New Jersey 08540, USA}

\author{Giulio Corazza}
\affiliation{Department of Physics, ETH Z\"{u}rich, Z\"{u}rich 8093, Switzerland}


\author{Yiwen Chu}
\affiliation{Department of Physics, ETH Z\"{u}rich, Z\"{u}rich 8093, Switzerland}

\author{Matteo Fadel}
\email{matteofadel@pku.edu.cn}
\affiliation{Institute of Quantum Science and Technology $\&$ Frontiers Science Center for Nano-Optoelectronics, School of Physics, Peking University, Beijing 100871, China}
\affiliation{Department of Physics, ETH Z\"{u}rich, Z\"{u}rich 8093, Switzerland}

\date{\today}

\begin{abstract}
Massive mechanical oscillators couple directly to forces, accelerations and gravity, making them attractive probes of new physics and of quantum mechanics at macroscopic scales.
Pushing sensitivities beyond the standard quantum limit for signals of unknown phase requires non-Gaussian quantum states, whose metrological advantage has so far been confined to lighter systems such as trapped ions.
Here we operate a microgram-scale acoustic mode of a high-overtone bulk acoustic wave resonator as a phase-insensitive quantum-enhanced sensor.
A transmon qubit prepares the mode in Fock states of up to seven phonons by optimal control, and resolves its phonon-number distribution after interaction with a weak resonant signal.
Bayesian analysis yields a Fisher information five times the vacuum value, a \unit{7.0}{dB} gain in signal-amplitude sensitivity over any classical probe state, irrespective of its energy, currently limited by state preparation infidelities.
We then train a neural network on the measured sensor response to perform Bayesian estimation without any analytical model of the system, showing the same sensitivity gain with half the systematic bias.
Our results establish acoustic resonators prepared in non-Gaussian states as quantum-enhanced probes for high-frequency gravitational waves, dark matter and wave-function collapse.
\end{abstract}

\maketitle

\makeatletter
\let\oldaddcontentsline\addcontentsline
\renewcommand{\addcontentsline}[3]{}
\makeatother

Quantum fluctuations impose a fundamental noise floor on measurement sensitivities, limiting uncorrelated sensing strategies to the standard quantum limit (SQL).
Overcoming this bound is a central objective of quantum metrology, which harnesses non-classical states to push sensitivity toward the Heisenberg limit \cite{caves_quantum-mechanical_1981, giovannetti_quantum-enhanced_2004, giovannetti_advances_2011}.
The demonstration of such quantum-enhanced strategies has been particularly successful in continuous-variable (CV) systems \cite{fadel_quantum_2025}, such as optical fields \cite{abadie_gravitational_2011, aasi_enhanced_2013, tse_quantum-enhanced_2019, acernese_increasing_2019}, microwave resonators \cite{wang_heisenberg-limited_2019, wang_quantum-enhanced_2022, deng_quantum-enhanced_2024, pan_realization_2025}, and mechanical degrees of freedom \cite{mccormick_quantum-enhanced_2019, wolf_motional_2019,rahman_genuine_2025, gilmore_quantum-enhanced_2021, mason_continuous_2019}, where states in the unbounded Hilbert space of a quantum harmonic oscillator can be exploited to accurately estimate perturbations that act as e.g. phase-space displacements and rotations.
Paradigmatic examples include the preparation of squeezed \cite{pirkkalainen_squeezing_2015,lecocq_quantum_2015, wollman_quantum_2015, burd_quantum_2019, marti_quantum_2024}, cat \cite{monroe_schrodinger_1996,vlastakis_deterministically_2013,bild_schrodinger_2023}, and Gottesman--Kitaev--Preskill states \cite{gottesman_encoding_2001, fluhmann_encoding_2019, duivenvoorden_single-mode_2017,campagne-ibarcq_quantum_2020,valahu_quantum-enhanced_2025}, each offering distinct metrological advantages depending on the specific sensing task \cite{fadel_quantum_2025}.
Among these, Fock states are particularly attractive for displacement sensing: their rotational symmetry in phase space makes them optimal when the displacement direction is unknown or randomly distributed, saturating the energy-constrained bound on the quantum Fisher information irrespective of the signal phase \cite{fadel_quantum_2025, wolf_motional_2019, rahman_genuine_2025, deng_quantum-enhanced_2024}.

Although CV quantum metrology with light plays a crucial role in enhancing the sensitivity of, e.g., optical interferometers, applying these techniques to mechanical degrees of freedom, such as the motional states of trapped ions or solid-state mechanical oscillators, extends quantum-enhanced sensing to systems with a finite mass. 
This is of great interest, since massive probes are directly sensitive to weak forces \cite{biercuk_ultrasensitive_2010, monteiro_force_2020}, accelerations \cite{krause_high-resolution_2012, zhou_broadband_2021}, gravity \cite{westphal_measurement_2021,middlemiss_measurement_2016}, and potential signatures of new physics \cite{carney_mechanical_2021,linehan_listening_2025,omahen_ultracold_2026}, including mechanisms that may induce wave-function collapse in macroscopic objects \cite{bassi_models_2013, vinante_narrowing_2020, gely_superconducting_2021}. 
Still, the coherent preparation, control, and detection of quantum states of motion in such systems pose significant experimental challenges.
In particular, while sensing protocols beating the SQL with squeezed states of motion have been demonstrated, their advantage is restricted to a known quadrature, and the phase-insensitive enhancement offered by non-Gaussian probes remains to be demonstrated in a massive mechanical system.

In recent years, the emergence of circuit quantum acoustodynamics (cQAD) has enabled exquisite control over the phononic degrees of freedom of high-frequency mechanical resonators by coupling them to superconducting qubits \cite{oconnell_quantum_2010, gustafsson_propagating_2014, chu_quantum_2017, manenti_circuit_2017, satzinger_quantum_2018}. 
In this context, high-overtone bulk acoustic wave resonators (HBARs) confine vibrations within a solid-state substrate.
The resulting mechanical modes have large effective masses, typically in the microgram range \cite{bild_schrodinger_2023}, which makes them well suited to probing the boundaries of quantum mechanics and searching for new physics. 
Concrete applications range from the detection of gravity-related effects, such as spacetime modifications \cite{marin_gravitational_2013,campbell_improved_2023,donadi_quantum_2025}, high-frequency gravitational waves \cite{goryachev_gravitational_2014, goryachev_rare_2021,omahen_ultracold_2026}, and dark matter candidates \cite{campbell_searching_2021, linehan_listening_2025}, to placing empirical bounds on collapse models that propose nonlinear modifications to the Schr\"odinger equation \cite{schrinski_macroscopic_2023}.

\begin{figure*}[t!]
    \centering
    \includegraphics[width=\textwidth]{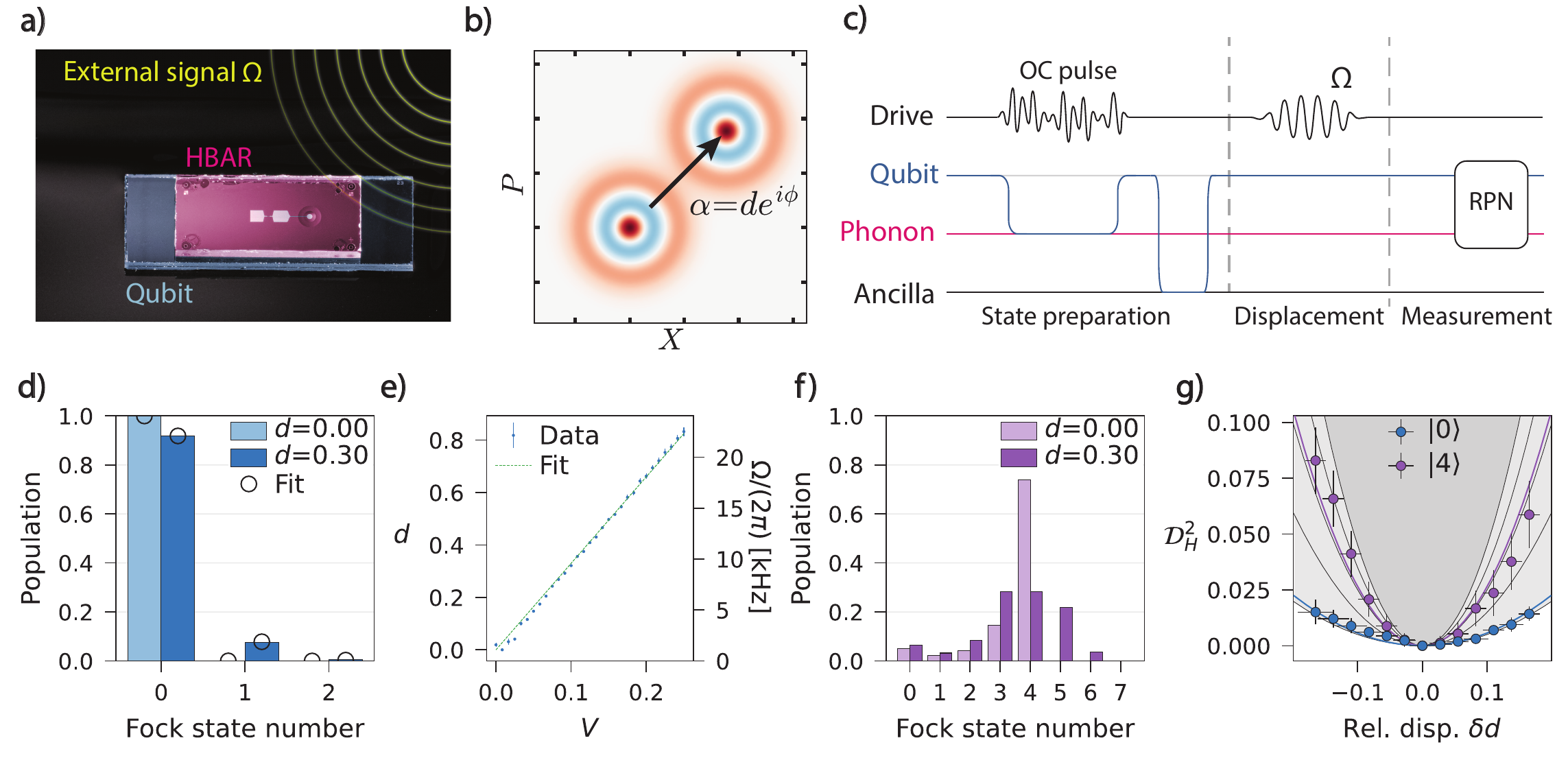}
    \caption{\textbf{Mechanical resonator device and sensing protocol.} 
    \textbf{(a)} False-colour image of the cQAD device, comprising a superconducting transmon qubit (blue) coupled to a high-overtone bulk acoustic wave resonator (HBAR, pink). An external signal of amplitude $\Omega$ is sensed by the device.
    \textbf{(b)} Illustration of displacement sensing in phase space. The external drive displaces the Wigner distribution of the prepared Fock-state probe by an amount $\alpha$, indicated by the arrow.
    \textbf{(c)} Experimental sequence for the sensing protocol. A shaped optimal control (OC) pulse prepares the target Fock state while the qubit is tuned into resonance with the phonon mode; the qubit is subsequently reset by using an ancilla mode. The signal is encoded by applying a resonant drive of amplitude $\Omega$ to displace the phonon state, followed by resonant phonon-number (RPN) readout.
    \textbf{(d)} Measured phonon-number populations for the vacuum probe after displacements of amplitude $d=0$ and $0.30$. Open circles show Poissonian fits used to calibrate the displacement amplitude.
    \textbf{(e)} Calibration of the displacement-drive amplitude $\overline{\Omega}$ (left axis) and corresponding phase-space displacement $d$ (right axis) as a function of the applied voltage $V$. The dashed line is a linear fit to the data.
    \textbf{(f)} Measured phonon-number populations for a probe prepared in $\lvert4\rangle$, before ($d=0$) and after ($d=0.30$) the displacement.
    \textbf{(g)} Squared Hellinger distance $\sqHell$ between phonon-number distributions separated by a displacement $\delta d$, for probes prepared in $\lvert0\rangle$ and $\lvert4\rangle$. Solid curves are quadratic fits around $\delta d=0$, whose curvature determines the Fisher information through $\sqHell=F\,\delta d^2/8$. Black lines are for $F=4(2k+1)$ with $k\in\{0,\dots,4\}$ from bottom to top.
    Shaded gray area indicates sensitivities inaccessible by classical probe states and the dark gray area is inaccessible by any probe with $\bar{n}\leq 4$.
    }
    \label{fig:1}
\end{figure*}

Here, we demonstrate the phase-insensitive quantum-enhanced sensing of the amplitude of a weak signal using an HBAR oscillator prepared in Fock states of motion.
Optimal-control pulses initialize the acoustic mode in states up to $\ket{7}$, after which a weak resonant drive imprints a small perturbation corresponding to a phase-space displacement of unknown phase, read out by a phonon-number resolving measurement.
Independent analyses of the measured distributions agree on a Fisher information five times the value accessible to any classical state, irrespective of its energy, corresponding to a $\unit{7.0}{dB}$ gain in sensitivity beyond the standard quantum limit.
The enhancement we observe does not grow indefinitely with the probe energy, but peaks for the state prepared in $\ket{4}$, beyond which imperfect state preparation outweighs the ideal Fock-state gain.
The estimation above relies on a parametrized model of the device response, whose residual mismatch enters the estimate as a systematic error.
We therefore replace it by a model-free estimator, a neural network trained directly on the calibration records, which reproduces the same quantum advantage while halving the systematic bias.
Since the estimator is learned from data alone, the inference becomes independent of any analytical description of the device, a prerequisite for sensing with probes whose response is not known in analytical form.
Together, these results establish quantum acoustics as a platform for quantum-enhanced metrology with massive oscillators, and provide a roadmap towards their use for exploring new physics.

\section{Experimental platform}

The cQAD device used in this work consists of a transmon qubit flip-chip bonded to an HBAR \cite{marti_quantum_2024}, see Fig.~\ref{fig:1}a.
The qubit has an idle frequency of $\omega_q/2\pi = \unit{5.074}{GHz}$, which can be tuned via a far-off-resonant Stark shift drive. At this frequency, the qubit has an energy relaxation time $T_{1,q}=\unit{27(1)}{\micro\second}$, a Ramsey decoherence time $T_{2,q}^\ast=\unit{18(1)}{\micro\second}$, and an anharmonicity $\alpha_q/2\pi=\unit{185}{MHz}$.
The coupling between qubit and HBAR is mediated by a piezoelectric transducer made of aluminum nitride and it takes the form of a Jaynes-Cummings (JC) interaction with coupling strength $g/2\pi=\unit{269}{kHz}$.
The phonon mode we consider in this work has a frequency $\omega_a/2\pi=\unit{5.088}{GHz}$, an energy relaxation time $T_{1,a}=\unit{196(8)}{\micro\second}$, a Ramsey decoherence time $T_{2,a}^\ast=\unit{401(4)}{\micro\second}$, and an effective mass of about $\unit{16}{\mu g}$ \cite{bild_schrodinger_2023}.
Our system can be described by the Hamiltonian
\begin{align}
    \hat{H}_{\text{cQAD}}/\hbar &= \omega_q \hat{q}^\dagger \hat{q} - \dfrac{\alpha_q}{2} \hat{q}^{\dagger\,2} \hat{q}^2 \notag\\
    &\quad + \omega_a \aad \aa + g(\hat{q} \aad + \hat{q}^\dagger \aa) + \hat{H}_{\text{qd}}/\hbar \;, \label{eq:fullH}
\end{align}
where $\hat{q}$ and $\hat{a}$ are the bosonic annihilation operators for the qubit and the phonon mode, respectively. The term $\hat{H}_{\text{qd}}/\hbar=\Omega_d(t) e^{-i\omega_d t} \hat{q}^\dagger + \text{h.c.}$ describes a microwave drive at frequency $\omega_{d}$ and time-dependent complex amplitude $\Omega_d(t)$ applied to the qubit. 
This drive will be used for the preparation of the initial sensing state, as well as to generate the phase space displacement to be estimated.

For the first task, namely the state preparation, we tune the qubit frequency to operate at $\omega_q=\omega_a=\omega_d$ and apply a drive $\Omega_d(t)$ determined using quantum-optimal-control theory in order to realize the mapping $\ket{g}\ket{0} \rightarrow \ket{g}\ket{k}$. This brings the original vacuum state of the phonon mode $\ket{0}$ into Fock state $\ket{k}$, while leaving the qubit in its ground state $\ket{g}$. 
Because of decoherence, pulse distortion, and other experimental imperfections, we see that Fock states with $k\in\{1,\dots,7\}$ are prepared with fidelities $\mathcal{F}_k=\{93.6\%, 95.0\%, 94.0\%, 92.0\%, 79.9\%, 76.2\%, 67.4\% \}$ \cite{sm}. 

For the second task, namely emulation of the signal to be sensed, we operate instead with the bare $\omega_q$ and $\omega_a$, and set $\omega_d=\omega_a$. 
Since our $\Delta=\omega_a-\omega_q$ gives $|g/\Delta|\ll1$ and $|\Omega_d/\Delta|\ll1$, the qubit follows the drive adiabatically and can be eliminated.
In the rotating frame of the phonon mode, this results in the effective Hamiltonian
\begin{equation}\label{eq:dispH}
  \hat H/\hbar = i \Omega(t) \left(e^{i\phi}\aad-e^{-i\phi}\aa\right) \fcomma
\end{equation}
where $\Omega(t)\simeq|g/\Delta|\,|\Omega_d(t)|$ is the real envelope of the drive as seen by the mode and $\phi$ the resulting phase-space displacement direction, fixed within each shot.
Because $\hat H(t)$ commutes with itself at different times, the evolution is the displacement $\hat D(\alpha(t))$ with $\alpha(t)=d(t)\,e^{i\phi}$ and amplitude $d(t)=\int_{0}^{t}d\tau\,\Omega(\tau)$.
In practice, however, phonon losses lead to a reduction of this amplitude, as we discuss in detail at the end of the next section.

\section{Displacement sensing theory}


Our goal is to estimate the unknown amplitude $d$ of a phase-space displacement with quantum-enhanced sensitivity.
In a parameter-estimation task, the quantity $d$ is encoded into the probe state $\rho_d$, and read out by a measurement described by a positive operator-valued measure (POVM) $\{\hat E_\mu\}_\mu$, whose outcomes occur with probabilities $P(\mu|d)=\Tr[\hat E_\mu\rho_d]$.
From $M$ independent repetitions, an estimator $\Theta$ of the parameter is constructed.
If $\Theta$ is unbiased in a neighbourhood of $d$, its variance obeys the classical Cram\'er--Rao bound (CRB)
\begin{equation}\label{eq:CCRB}
    \var{\Theta} \geq \dfrac{1}{M\,F[d]} \fcomma \qquad
    F[d] = \sum_\mu \dfrac{1}{P(\mu|d)} \left(\dfrac{\partial P(\mu|d)}{\partial d}\right)^{2} \fcomma
\end{equation}
where $F[d]$ is the single-shot classical Fisher information (FI), which quantifies the metrological performance of the chosen probe state and measurement.
Maximizing $F[d]$ over all POVMs defines the quantum Fisher information (QFI) $F_Q[d]$, which is thus a property of the probe state only.
Since $F[d]\leq F_Q[d]$ for every measurement, Eq.~\eqref{eq:CCRB} implies the quantum Cram\'er--Rao bound (QCRB) $\var{\Theta}\geq 1/(M\,F_Q[d])$.
For a unitary encoding $\rho_d=e^{-id\hat G}\rho\,e^{id\hat G}$ with Hermitian generator $\hat G$ and a pure probe state $\rho=\ket{\psi}\bra{\psi}$, the QFI reduces to $F_Q[d]=4\var{\hat G}_{\ket{\psi}}$, which is independent of $d$ because the encoding leaves all moments of $\hat G$ invariant.

In the case of the displacement generated by Eq.~\eqref{eq:dispH}, the encoding is unitary with $\hat G(\phi) = i(e^{i\phi}\aad-e^{-i\phi}\aa)$. 
When the direction $\phi$ is unknown or fluctuates from shot to shot a relevant figure of merit is the worst-case QFI over the displacement direction, which for any state with mean phonon number $\bar{n}$ obeys \cite{fadel_quantum_2025}
\begin{equation}\label{eq:avQFI}
    \min_\phi F_Q[d,\{\rho,\hat G(\phi)\}] \leq 4(2\bar{n}+1) \fperiod
\end{equation}
Fock states $\ket{k}$ saturate this bound, since $F_Q[d,\{\ket{k},\hat G(\phi)\}]=4(2k+1)$ for every $\phi$ and $d$ \cite{sm}.
The $\phi$-independence reflects the rotational symmetry of the Fock-state Wigner function in phase space: unlike phase-sensitive probes such as squeezed states, Fock states respond identically to a displacement in any direction.
Coherent states are likewise isotropic, but give $F_Q[d]=4$ irrespective of their energy, so that Fock states are optimal probes for this task at fixed mean phonon number.
We further show in \cite{sm} that, for a pure Fock state and in the absence of losses, phonon-number measurements attain the QFI for estimating the displacement amplitude $d$, which motivates the resonant phonon-number (RPN) readout used in our experiments.
The value $F_Q[d]=4$ is in fact a bound for all classical probes: since the QFI is convex, any state that can be written as a mixture of coherent states (i.e. with a positive Glauber-Sudarshan $P$ function) satisfies $F\leq F_Q\leq4$.
Therefore, besides certifying non-classicality, violating this bound witnesses metrologically useful quantum resources and sub-shot-noise sensitivity.

In a realistic scenario, the probe system is inevitably coupled to its environment, so that the actual sensitivity falls below the ideal value of Eq.~\eqref{eq:avQFI}.
In the following we consider coupling to a zero-temperature bath at rate $\Gamma=1/T_{1,a}$, where $T_{1,a}$ is the energy relaxation time of the mode, such that the probe state evolution during the displacement is governed by the master equation
\begin{equation}
  \dot\rho
  = -\frac{i}{\hbar}[\hat H,\rho]
    + \Gamma\left(\hat a\rho\hat a^\dagger
      -\frac{1}{2}\{\hat a^\dagger\hat a,\rho\}\right) \;.
  \label{eq:mastereq}
\end{equation}
This dynamics defines a Gaussian channel that decomposes exactly into a pure-loss channel of transmissivity $\eta=e^{-\Gamma t}$, followed by a unitary displacement of the attenuated amplitude $d(t)=\int_0^t d\tau\,e^{-\Gamma(t-\tau)/2}\,\Omega(\tau)$ \cite{sm}.
Since $\eta$ does not depend on the drive, the encoding of $d$ remains unitary, but it now acts on a probe state already degraded by losses.
For an initial coherent state the degraded state is again coherent, so that the classical benchmark $F_Q=4$ is unaffected by loss.
An initial Fock state $\ket{k}$, by contrast, is mapped onto a Fock-diagonal mixture, and its QFI must be evaluated from the general mixed-state expression \cite{sm}, giving $F_Q[d]<4(2k+1)$ for any $\eta<1$.
The sensitivity of Fock states grows linearly with their phonon number, but at the same time also state preparation infidelity and loss becomes more severe.
The competition between these effects actually leads to an optimal phonon number, which we later identify.

\section{Experimental protocol}

The protocol we follow for our measurements is illustrated in Fig.~\ref{fig:1}c.
At the beginning of every experimental run, the phonon mode to be used for sensing is, to very good approximation, in the vacuum state $\ket{0}$ \cite{omahen_ultracold_2026}.
From this initial condition, we may prepare the mode in higher Fock states $\ket{k}$ through optimal-control strategies.
The qubit state is then reset by swapping any residual population into an ancillary phonon mode \cite{omahen_reset_2026}.
This concludes the preparation of our quantum sensor, which can now be used to probe an external perturbation.

The weak signal to be sensed is emulated by a microwave drive pulse on resonance with the phonon mode, realizing the Hamiltonian Eq.~\eqref{eq:dispH}.
In each experimental run the drive is applied for a time $t_d$ at a constant microwave generator voltage setting, selected from a discrete set $\{V_i\}_{i=1}^{N_d}$. 
Each $V_i$ corresponds to an effective drive amplitude $\Omega_i$ on the phonon mode and hence a displacement $d_i$, which we calibrate using the vacuum as a probe.
Concretely, for each $V_i$, the displacement pulse is applied to $\ket{0}$ and the resulting phonon-number distribution $P(\mu|V_i)$ is reconstructed by RPN measurement, see e.g.\ Fig.~\ref{fig:1}d.
Since the vacuum remains a coherent state under Eq.~\eqref{eq:mastereq}, these distributions are Poissonian with mean $\bar n=d^2$, so that a Poissonian fit to each of them yields the displacement amplitude associated with each voltage.
As these two are directly proportional, a linear fit of the extracted $d_i(V_i)\sim\sqrt{\bar n(V_i)}$ provides the experimental conversion $V\mapsto d(V)$ used to define the calibrated displacement grid $\{d_i\}_{i=1}^{N_d}$ \cite{sm}.
We have $N_d=31$ points between $d_1=0$ and $d_{31}=0.832$, see Fig.~\ref{fig:1}e.

The displacement pulse consists of a constant-amplitude section of duration $t_c=\unit{5.0}{\micro\second}$ and two smooth ramps of duration $\timeedge=\unit{0.5}{\micro\second}$, giving a total duration $t_d=\unit{6.0}{\micro\second}$ and an effective square-pulse duration $\timedispeff=t_c+\timeedge=\unit{5.5}{\micro\second}$.
Replacing the ramped pulse by a constant drive of the same integrated area over the total duration $t_d$, which is an excellent approximation for $\timeedge\ll t_d$ and $\Gamma\timeedge\ll1$, the attenuated amplitude introduced above evaluates to \cite{sm}
\begin{equation}\label{eq:dToOmega}
    d = \frac{2\Omega}{\Gamma} \dfrac{\timedispeff}{t_d} \left(1 - e^{-\Gamma t_d /2} \right) \;,
\end{equation}
which can be used to convert a measured displacement $d$ into the corresponding drive amplitude $\Omega$.
In the lossless limit, $\Gamma t_d\rightarrow0$, this reduces to $d=\Omega\,\timedispeff$.
In the rest of this work we will focus on the precise estimation of $d$.

The monotonic dependence of the mean phonon number on the displacement amplitude can already be used to estimate an unknown $d$ from the measured $\bar{n}(d)$, e.g.\ through the method of moments \cite{fadel_quantum_2025}.
The full phonon-number distribution $P(\mu|d)$, however, contains additional information that can be used to reduce the estimation uncertainty.
One possibility is maximum likelihood estimation, which saturates the CRB asymptotically in the number of measurements.
For scarce data, on the other hand, Bayesian analysis is often favored.
Among its advantages is the fact that it requires no minimum sample size while remaining asymptotically optimal, it naturally incorporates prior information, and it directly provides meaningful confidence intervals.

\begin{figure*}
    \centering
    \includegraphics[width=0.98\textwidth]{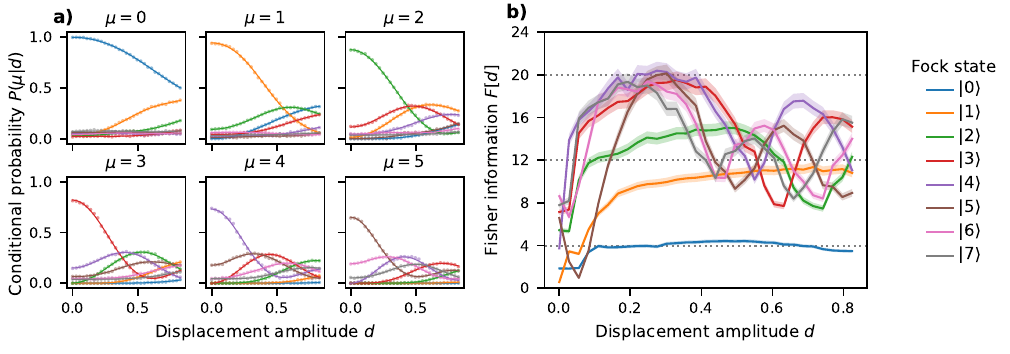}
    \caption{\textbf{Displacement response and Fisher information of Fock-state probes.} 
    \textbf{(a)}, Conditional probabilities $P(\mu|d)$ for detecting $\mu\in\{0,\ldots,5\}$ phonons as a function of the displacement amplitude $d$, for an initial Fock state $\lvert k\rangle$ with $k\in\{0,\ldots,7\}$. Points show the experimentally measured probabilities and solid curves the fits to the analytical model. 
    \textbf{(b)}, Fisher information $F[d]$ obtained from the fitted phonon-number distributions in \textbf{a}, for the same initial Fock states. Dotted horizontal lines indicate the ideal Fock-state quantum Fisher information $F_Q[d]=4(2k+1)$ for $k=0,1,2$. Shaded regions indicate the uncertainty of the inferred Fisher information.
    }
    \label{fig:2}
\end{figure*}

\section{Fisher information from the Hellinger distance}

For a given probe state, a direct estimate of its sensitivity to displacements under phonon-number measurements can be obtained from the measured $P(\mu|d)$.
The method is based on the squared Hellinger distance between the distributions at two nearby displacements,
\begin{equation}\label{eq:sqHdist}
    \sqHell(d,\delta d) = 1 - \sum_{\mu} \sqrt{P(\mu|d) P(\mu|d + \delta d)} \fcomma
\end{equation}
which is half the squared Euclidean distance between the corresponding probability amplitudes $\sqrt{P(\mu|d)}$, and whose square root is a metric on the space of probability distributions \cite{wootters_statistical_1981, braunstein_statistical_1994}.
Crucially, $\sqHell$ vanishes quadratically in $\delta d$ with a curvature set by the Fisher information,
\begin{equation}\label{eq:Hellexp}
    \sqHell(d,\delta d) = \frac{F[d]}{8}\,\delta d^2 + \mathcal{O}(\delta d^3) \fperiod
\end{equation}

In our experiment we measure $P(\mu|d_i)$ on the calibrated grid $\{d_i\}_{i=1}^{N_d}$ and evaluate Eq.~\eqref{eq:sqHdist} for pairs of displacements separated by small shifts $\delta d=d_j-d_i$ around the chosen point $d_i$.
Because of the relation Eq.~\eqref{eq:Hellexp}, we then extract $F[d_i]$ from a quadratic fit centered at $\delta d = 0$.
An example for probe states $\ket{0}$ and $\ket{4}$ is shown in Fig.~\ref{fig:1}d,f,g, from which we infer $F=21.2(1.3)$ for $\ket{4}$ at $d\simeq0.3$, thus witnessing non-classical sensitivity.

The same measurements of $P(\mu|d)$ also provide the calibration needed to estimate an unknown displacement $d'$ from a finite record of single-shot outcomes acquired at $d'$.
A powerful approach for this task, particularly convenient when the measurement statistics are scarce, is Bayesian inference, which we discuss in the next section.

\section{Bayesian estimation}

To demonstrate the Bayesian estimation protocol we start from the measured conditional probabilities $P(\mu|d_i)$, shown in Fig.~\ref{fig:2}a for different initial Fock states, which constitute the calibration of the sensor.
Using these sampled probabilities directly as a likelihood is problematic for low-probability outcomes.
Finite sampling produces large relative fluctuations and, for rare occasions, zeros, which are amplified when many likelihood factors are multiplied and can artificially suppress parts of the posterior.
We therefore fit the measured probabilities with a smooth model motivated by the analytical displacement matrix elements obtained from Eq.~\eqref{eq:mastereq} \cite{sm},
\begin{equation}\label{eq:effectiveFit}
P_\text{fit}(\mu|d)=\sum_{m=0}^{\dim[\mathcal H]-1} W_m^{(\mu)}\left|\langle\mu|\hat D(d)|m\rangle\right|^2 \fcomma
\end{equation}
which gives a continuous estimate of the response throughout the calibrated range, removing the restriction to the discrete experimental grid.

The matrix elements $|\langle\mu|\hat D(d)|m\rangle|^2$ are those of an ideal displaced Fock state, so that for a Fock-diagonal probe and perfect readout the coefficients $W_m^{(\mu)}$ would reduce to the $\mu$-independent populations of the prepared state.
Letting them depend on the outcome $\mu$ instead, and treating them as free fit parameters, phenomenologically accounts for imperfections in state preparation, evolution and readout \cite{sm}.
The truncation dimension $\dim[\mathcal H]$ is chosen sufficiently large that neglected higher-phonon-number populations have a negligible effect over the full fitted displacement range.
As shown in Fig.~\ref{fig:2}a, the resulting model accurately follows the measured data.

For a record of $M$ independent phonon-number outcomes $\vect{\mu}=\{\mu_i\}_{i=1}^{M}$, all acquired at the same unknown displacement, Bayes' theorem gives
\begin{equation}\label{eq:bayesfit}
P_\text{fit}(d|\vect{\mu})\propto P(d)\prod_{i=1}^{M}P_\text{fit}(\mu_i|d) \fcomma
\end{equation}
where $P(d)$ is the prior distribution.
We take it uniform over the calibrated interval, so that the posterior is determined by the record and by the fitted response alone.

For sufficiently large $M$, and away from the boundaries of the parameter interval, the posterior becomes asymptotically Gaussian, $P_\text{fit}(d|\vect{\mu})\simeq \exp[-(d-\Theta(\vect{\mu}))^2/2\sigma^2]/\sqrt{2\pi\sigma^2}$, where the posterior mean $\Theta(\vect{\mu})$ provides a point estimate of the displacement amplitude and $\sigma$ quantifies its uncertainty.
In this regime the Bernstein--von Mises theorem gives $\sigma^2\simeq[M F[d_0]]^{-1}$ for data generated at the true displacement $d_0$, so that the single-shot Fisher information can be inferred as $F[d_0]\simeq1/(M\sigma^2)$.
Strictly, this returns the Fisher information of the fitted response, which coincides with that of the device as long as Eq.~\eqref{eq:effectiveFit} describes the data faithfully. 
Repeating the procedure across the calibrated grid yields the Fisher information shown in Fig.~\ref{fig:2}b.
The largest values, $F\simeq20$ around $d\simeq0.3$, are reached by the probes prepared in $\ket{k}$ with $3\leq k\leq7$, consistent with the value $F=21.2(1.3)$ obtained independently for $\ket{4}$ from the Hellinger analysis.

\begin{figure}[t!]
    \centering
    \includegraphics[width=0.95\linewidth]{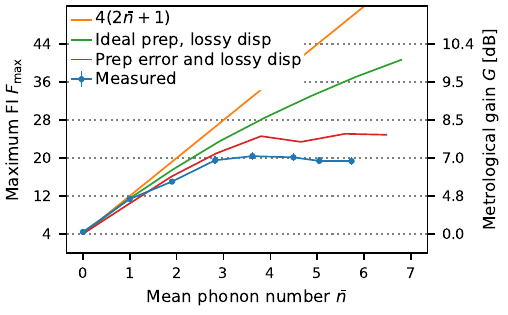}
    \caption{\textbf{Sensitivity scaling.} 
    Maximum Fisher information $F_{\max}$ as a function of the (final) mean phonon number $\bar{n}$ of the probe state.
    The orange line denotes the ideal Fisher information scaling $F_Q=4(2\bar{n}+1)$, achieved by Fock states and unitary displacements.
    The green curve is the $F_Q$ achieved by Fock states in the presence of losses during the displacement, as described by Eq.~\eqref{eq:mastereq} for our experimental parameters.
    The red curve further includes state preparation errors, as described by an ansatz state of the same fidelity $\mathcal{F}_k$ as the one measured \cite{sm}.
    Blue points show the Fisher information measured experimentally, which saturates for $\bar{n}\gtrsim 4$.
    This saturation arises from the competition between the metrological gain of higher-order Fock states and losses.
    The gap between data (blue) and model (red) is attributed to finite measurement resolution, parameter uncertainties, and to the fact that for mixed states phonon-number measurements give $F\leq F_Q$.
    }
    \label{fig:3}
\end{figure}

\section{Quantum advantage}

We now investigate how the experimentally observed sensitivity scales with the resources of the probe.
We repeat the Bayesian analysis for the different prepared Fock states and characterize each probe by the maximum Fisher information $F_{\max}=\max_d F[d]$ reached over the measured displacement range.
Since the experimentally prepared states are not ideal Fock states, we quantify their energy from their mean phonon number $\bar{n}$ rather than by the target Fock-state index $k$.
The resulting $F_{\max}$ is shown as a function of $\bar{n}$ in Fig.~\ref{fig:3}.

In the ideal lossless case, the maximum QFI for phase-insensitive displacement sensing increases linearly with the mean phonon number as $4(2\bar{n}+1)$, which can be achieved by phonon-number measurements.
The experimental values initially follow this increase, with all excited probes exceeding the classical benchmark $F_Q=4$.
At larger $\bar{n}$, however, the gain saturates: $F_{\max}$ reaches its largest value for the probe prepared in $\ket{4}$ and then remains close to $F\simeq20$ for the higher Fock states, despite the larger energy stored in the oscillator.
Increasing the phonon number therefore does not indefinitely improve the experimentally accessible sensitivity.
This behavior reflects the competing roles of energy, decoherence and preparation fidelity, as we illustrate in Fig.~\ref{fig:3} by comparing the data to two reference models \cite{sm}.

The first (green curve) assumes an ideal Fock state $\ket{k}$ subject to phonon loss during the sensing sequence, as described by Eq.~\eqref{eq:mastereq}.
Loss alone, however, does not reproduce the observed saturation.
This model does lead to a plateau, at the value $2/(1-\eta)$, but only once the mean number of lost phonons $k(1-\eta) \gg 1$ \cite{sm}.
For our parameters, $\eta\simeq 0.97$, which gives a saturation at $F_Q\simeq 67$ for $k\gg 50$, so that over the measured range $k\leq7$ the green curve is still rising, though already sublinearly.
With the available $T_{1,a}$, a perfectly prepared Fock state would therefore continue to gain sensitivity well beyond the occupations explored here.

The observed saturation can be explained by taking into account our imperfect state preparation.
The measured zero-displacement fidelities $\mathcal{F}_k$ drop with $k$, so that the higher-energy probes are mixtures of Fock states.
The second reference model (red curve) accounts for this through a minimal ansatz assigning the $1-\mathcal{F}_k$ population to the state $\ket{k-1}$, as expected when the dominant preparation error is energy relaxation, and propagating the resulting Fock-diagonal state through the same lossy displacement channel \cite{sm}.
This yields a saturation at $F_Q\simeq 25$ for $\bar{n}\gtrsim 4$, in good agreement with the measurements.
The residual difference between the red curve and the measured $F_{\max}$ reflects the finite resolution of the RPN readout, parameter uncertainties, and the fact that, once the probe state is mixed, phonon-number measurements do not necessarily saturate the QFI \cite{sm}.

The observed $F_{\max}\simeq20$ is five times the value $F_Q=4$ attainable by any mixture of coherent states, so that the sensitivities reported here are inaccessible to classical probes of arbitrary energy. 
Since an efficient estimator reaches $\var{\Theta}=1/(MF)$, this amounts to a metrological gain $G=10\log_{10}(F_{\max}/4)\simeq\unit{7.0}{dB}$ over a vacuum probe subject to the same evolution and readout scheme, equivalently a factor $\sqrt{F_{\max}/4}\simeq2.2$ in displacement uncertainty reduction at fixed $M$. 
The Fisher information enters the force sensitivity as $S_F\propto F^{-1/2}$, with the remaining factors set by the pulse and the duty cycle alone, so the same gain applies directly to the smallest resolvable force and acceleration at equal averaging time.
This allows us to reach $\unit{0.77}{pN\,Hz^{-1/2}}$ for a $\unit{5}{GHz}$ force \cite{sm}.

\begin{figure*}
    \centering
    \includegraphics[width=\textwidth]{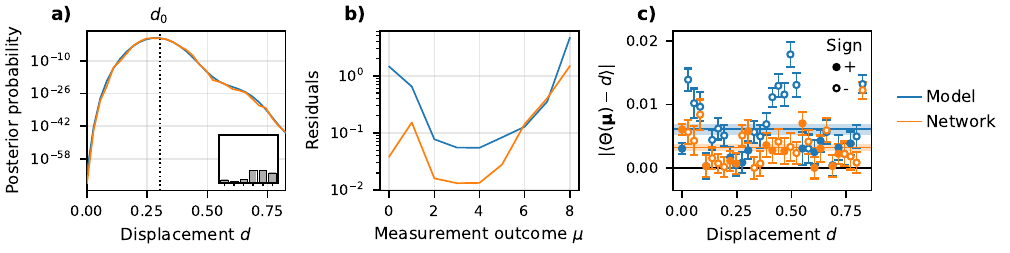}
    \caption{\textbf{Model-free displacement estimation with a neural network.} Comparison between the Bayesian analysis based on the fitted response Eq.~\eqref{eq:effectiveFit} (blue) and on the neural-network posterior Eq.~\eqref{eq:NNposterior} (orange), for a probe prepared in $\lvert 4\rangle$. \textbf{(a)} Posterior probability for a single record of $M=100$ single-shot outcomes acquired at the true displacement amplitude $d^\ast=0.30$ (dotted line). The two posteriors coincide over the full plotted range. Inset: histogram of the detected phonon numbers $\mu$ in this record. \textbf{(b)} Residuals between the measured posterior probabilities $P(d|\mu)$ and each reconstruction, as a function of the measurement outcome $\mu$ (summed over the calibrated displacement range). The network reproduces the data more accurately for the outcomes carrying most of the statistical weight. \textbf{(c)} Absolute estimator bias $|b(d)|=|\langle\Theta(\vect{\mu}) - d \rangle|$, Eq.~\eqref{eq:bias}. Each point is an average over 200 individual biases. Each bias has $M=100$ resampled records at each calibrated displacement. Filled and open symbols indicate positive and negative $b(d)$, error bars the standard error over records, and the horizontal bands the mean absolute bias over the full range, $3.2\times10^{-3}$ (network) and $6.1\times10^{-3}$ (fitted model).}
    \label{fig:4}
\end{figure*}

\section{Model-free estimation with a neural network}
\label{sec:NN}

A limitation of the Bayesian protocol described above is that it requires a continuous parametrization of the sampled response, Eq.~\eqref{eq:effectiveFit}. Although the model has a theoretical motivation, it still encodes assumptions on the working principle of the device, and its coefficients $W_m^{(\mu)}$ must absorb preparation, evolution and readout imperfections that are not independently characterized. Any mismatch between the model and the true response propagates into the posterior and, ultimately, into a systematic error of the estimated displacement.

This can be avoided with a black-box approach, in which the calibration data themselves define the estimator. We train a neural network (NN) to solve the classification task of assigning a single-shot phonon-number outcome $\mu$ to one of the $N_d$ calibrated displacements $\{d_j\}_{j=1}^{N_d}$, namely the map
\begin{equation}\label{eq:NNmap}
    f: \mu \;\longmapsto\; \vect{a} = \left(a_1,\ldots,a_{N_d}\right) \fcomma
\end{equation}
where $\vect{a}$ is the vector of output activations of the network. The input layer consists of a single neuron fed with the measured phonon number $\mu\in\{0,\ldots,\dim[\mathcal H]-1\}$, and the output layer of $N_d$ neurons with a softmax activation, so that $a_j\geq0$ and $\sum_j a_j=1$ by construction. Between them we use $n_\text{h}$ fully connected hidden layers with ReLU activations, whose depth and width are optimized depending on the initial Fock state $\ket{k}$ (e.g. Fig.~\ref{fig:4} uses $n_h=2$ with 20 neurons in each layer). The network is trained on the calibration records of Fig.~\ref{fig:2}a by minimizing the cross-entropy between $\vect{a}$ and the one-hot label of the displacement at which each shot was recorded, using for each displacement 80\% of the data for training and the remaining 20\% for validation.

Because the training set contains the same number of shots for every $d_j$, the softmax outputs approximate the single-shot posterior associated with a flat prior on the calibration grid,
\begin{equation}\label{eq:NNposterior}
    P_\text{NN}(d_j|\mu) = \dfrac{a_j}{\delta d} \fcomma \qquad
    \sum_{j=1}^{N_d} P_\text{NN}(d_j|\mu)\,\delta d = \sum_{j=1}^{N_d} a_j = 1 \fcomma
\end{equation}
with $\delta d=d_{j+1}-d_j$ the (constant) grid spacing. Note that the network returns the posterior directly: the conditional probabilities $P_\text{NN}(\mu|d)$ cannot be recovered from Eq.~\eqref{eq:NNposterior} without knowledge of the outcome distribution $P(\mu)$. The latter can nevertheless be obtained self-consistently by solving the implicit equation \cite{nolan_machine_2021}
\begin{align}\label{eq:NNprior}
    P_\text{NN}(d_j) &= \sum_\mu P_\text{NN}(d_j|\mu)\, P(\mu) \notag\\
    &= \sum_\mu P_\text{NN}(d_j|\mu) \sum_{k=1}^{N_d} P(\mu|d_k)\, P_\text{NN}(d_k)\,\delta d \fcomma
\end{align}
which we evaluate by iterating from a flat distribution until convergence. The resulting prior $P_\text{NN}(d_j)$ is consistent with the network response and with the measured outcome statistics. 
For our data, it agrees with a flat prior, confirming that the calibration grid is sampled uniformly enough not to bias the inference.

For a record of $M$ independent outcomes $\vect{\mu}=\{\mu_i\}_{i=1}^M$, Bayes' theorem combined with Eq.~\eqref{eq:NNposterior} gives
\begin{equation}\label{eq:NNmultishot}
    P_\text{NN}(d|\vect{\mu}) \;\propto\; P_\text{NN}(d) \prod_{i=1}^{M}
    \dfrac{P_\text{NN}(d|\mu_i)}{P_\text{NN}(d)} \fcomma
\end{equation}
where the division by $P_\text{NN}(d)$ removes the prior that is implicitly contained in each of the $M$ single-shot posteriors, and the $d$-independent factors $P(\mu_i)$ drop out upon normalization. From Eq.~\eqref{eq:NNmultishot} we extract the point estimate $\Theta(\vect{\mu})$ and the posterior width $\sigma$ exactly as in the model-based analysis, so that the two approaches can be compared on the same measurement records.

Figure~\ref{fig:4} summarizes this comparison. Panel a shows the posterior obtained from a single record of $M=100$ shots taken at the true displacement $d^\ast=0.30$, whose outcome histogram is displayed in the inset. Remarkably, the network posterior closely follows the one built on the fitted model of Eq.~\eqref{eq:effectiveFit}. In particular, both peak at $d^\ast$ and have the same width, meaning comparable Fisher information. 
The black-box estimator therefore reproduces the sensitivity reported in Figs.~\ref{fig:2}b and \ref{fig:3} without any assumption on the physical response of the device.

The two approaches differ, however, in how faithfully they describe the data. Fig.~\ref{fig:4}b shows the residuals between the measured posterior probabilities and each reconstruction, as a function of the detected phonon number $\mu$ (summed over the calibrated displacement range). The network reduces the residuals by up to an order of magnitude for the outcomes that carry most of the statistical weight, $2\lesssim\mu\lesssim6$, while both descriptions degrade for the rare outcomes at the edges of the distribution, where the calibration statistics are scarce. This is expected: with a single scalar input drawn from a finite alphabet, the network cannot be more expressive than a lookup table of conditional probabilities, and its role is not to extend the model space but to provide a regularized, assumption-free estimate of that table. It thus addresses the same problem that motivated the fit in the first place, namely the sampling zeros and large relative fluctuations of low-probability outcomes, without imposing a functional form.

The practical consequence appears in the estimator bias, which we calculate for an absolute maximum estimator on the experimental grid, Fig.~\ref{fig:4}c. 
For each calibrated displacement $d$ we resample records of $M$ shots from the measured data and evaluate
\begin{equation}\label{eq:bias}
    b(d) = \avg{\Theta(\vect{\mu}) - d } \fcomma
\end{equation}
where the average is over independent records.
While the fitted-model-based estimator exhibits significant bias, with systematic deviations around $d\simeq 0.5$, the network-based estimator appears unbiased within error bars across most displacements.
This is further illustrated by the fluctuations from point to point in the sign of $b(d)$ for the network, as expected for a purely statistical residual, whereas it appears more correlated over neighbouring displacements for the fitted model, as expected for a systematic error.
Averaged over the full calibrated range the mean absolute bias drops from $\overline{|b|}=6.1\times10^{-3}$ for the fitted model to $3.2\times10^{-3}$ for the network, i.e.\ by roughly a factor of two, and the residual value is compatible with the statistical floor set by the finite number of resampled records. 

This is the main advantage of the black-box approach. The Cramér--Rao bound of Eq.~\eqref{eq:CCRB} applies to unbiased estimators, so a systematic offset comparable to the posterior width would undermine the interpretation of $\sigma$ as a measure of the Fisher information. 
Keeping the bias below the statistical uncertainty, as achieved here, ensures that the quantum advantage reported in Fig.~\ref{fig:3} is not limited by data analysis. 
More generally, since the estimator is trained directly on calibration data, it applies to a device whose response is not known analytically, which is the case for many complex platforms.

\section{Conclusions and outlook}

We have demonstrated the operation of a microgram-mass acoustic mode of an HBAR device as a quantum-enhanced mechanical probe.
The mode is prepared in Fock states of up to seven phonons by optimal control, exposed to a weak resonant signal of random phase and unknown amplitude that displaces it in phase space, and read out by a phonon-number-resolving measurement.
Two independent analyses of the resulting phonon number distributions, based on the curvature of the squared Hellinger distance and on the width of the Bayesian posterior, agree on a Fisher information $F\simeq20$.
This is five times the maximum value attainable by any mixture of coherent states, meaning that the achieved sensitivity is inaccessible to classical probes of arbitrary energy.
A model-free neural-network estimator, trained directly on the calibration records to return the posterior for a single-shot outcome, reproduces the same value with a ``black-box'' approach that does not make any assumption on the device response.
Its residuals against the measured single-shot posterior are up to an order of magnitude smaller for the most frequent outcomes, and its systematic bias is roughly halved.
The largest gain is obtained for a probe prepared in $\ket{4}$, beyond which the quantum advantage saturates even though states of larger occupation could ideally be more sensitive.
This saturation is set neither by the energy relaxation time of the acoustic mode, which exceeds the protocol duration by more than an order of magnitude, nor by the resonant phonon-number detection, but by the state preparation fidelity at high phonon number.

Several directions can be explored to overcome current limitations.
A larger qubit-phonon coupling shortens the state preparation pulse, so that a device with $g/2\pi$ in the megahertz range would prepare the same states in a fraction of the present time and thus with higher fidelity.
Since the ideal Fock-state Fisher information grows linearly with the phonon number \cite{fadel_quantum_2025}, any fidelity improvement translates directly into a higher sensitivity, with no modification of the detection stage.
A second direction concerns the duty cycle: the time-normalized sensitivity is degraded by $\sqrt{t_\text{cycle}/t}$ \cite{gardner_stochastic_2026,sm}, and in our sequence the qubit and resonator reset dominates $t_\text{cycle}$, so that a faster conditional reset would improve the force sensitivity by $\sqrt{t_\text{cycle}/t_\text{cycle}'}$, i.e.\ by a factor of $\sim 12$ for the $\sim\unit{1}{\micro\second}$ reset times already demonstrated in circuit QED.
A third concerns the mode itself: since the zero-point fluctuation scales as $\xzpf\propto\omega^{-1/2}$ at fixed effective mass, operating a lower-frequency overtone of the same resonator improves the force and acceleration sensitivity as $\omega^{-1/2}$ \cite{sm}, e.g. by almost two orders of magnitude at \unit{1}{MHz}.

Taken together, these improvements would push the force and acceleration resolution of the present device far beyond what is reported here, into a regime relevant to searches for high-frequency gravitational waves \cite{goryachev_gravitational_2014, goryachev_rare_2021, linehan_listening_2025, omahen_ultracold_2026}, for dark-matter candidates \cite{campbell_searching_2021}, and for bounds on modifications to quantum mechanics \cite{bassi_models_2013, schrinski_macroscopic_2023, vinante_narrowing_2020}.

\vspace{3mm}
\textbf{Acknowledgments.--}
M.F. was supported by the Swiss National Science Foundation Ambizione Grant No. 208886, Spark Grant No. 237484, and The Branco Weiss Fellowship -- Society in Science, administered by the ETH Z\"{u}rich.

\clearpage
\newpage

\renewcommand{\addcontentsline}[3]{}
\bibliography{references}
\let\addcontentsline\oldaddcontentsline

\clearpage
\newpage

\renewcommand{\thetable}{S\arabic{table}}  
\renewcommand{\thepage}{\arabic{page}}  
\renewcommand{\thefigure}{S\arabic{figure}}
\renewcommand{\theHfigure}{S\arabic{figure}}
\renewcommand{\theequation}{S\arabic{equation}}
\setcounter{page}{1}
\setcounter{figure}{0}
\setcounter{table}{0}
\setcounter{section}{0}
\setcounter{equation}{0}

\widetext

{\centering\textbf{Supplementary Materials} \\}

\suppressfloats

\tableofcontents
\clearpage
\newpage

\section{Force and acceleration sensitivity}
\label{supp:resDrive}

Let us consider the Hamiltonian of a quantum harmonic oscillator subject to a time-dependent force
\begin{equation}\label{suppeq:driven_HO}
    \hat{H}_{\text{tot}} = \dfrac{\hat{p}^2}{2 \mathfrak{m}} + \dfrac{1}{2} \mathfrak{m}\, \omega^2 \hat{x}^2 - F(t) \hat{x} \fcomma
\end{equation}
where $\mathfrak{m}$ is the oscillator's mass, $\omega$ its frequency, and $F(t)$ the applied force.
Position and momentum operators are written in terms of the associated bosonic creation and annihilation operators as $\hat{x} = \sqrt{\hbar/2 \mathfrak{m}\omega}\, \left(\aa + \aad \right)$ and $\hat{p} = -i\sqrt{\hbar \mathfrak{m}\omega/2}\, \left(\aa - \aad \right)$.
Equation~\eqref{suppeq:driven_HO} can thus be written as
\begin{equation}
    \hat{H}_{\text{tot}} = \underbrace{\hbar\omega \left( \aad \aa + \dfrac{1}{2} \right)}_{\hat{H}_0} + \underbrace{\left[ - F(t) \xzpf \left(\aa + \aad \right) \right]}_{\hat{H}_F} \fcomma
\end{equation}
where $\xzpf \equiv \sqrt{\hbar/2 \mathfrak{m} \omega}$ is the zero-point fluctuation of the oscillator.

Consider now the case of a resonant driving force $F(t)=F_0 \sin(\omega t + \varphi) = \frac{F_0}{2i} \left( e^{i(\omega t +\varphi)} - e^{-i(\omega t +\varphi)} \right)$ with real amplitude $F_0 \geq 0$ and phase $\varphi$.
After moving into the interaction picture with respect to $\hat H_0$ using $\hat{U}_0=e^{-i \hat{H}_0 t /\hbar}$, we obtain
\begin{align}
\hat H &= \hat{U}_0^\dagger \hat{H}_F \hat{U}_0 \notag\\
&= \dfrac{iF_0 \xzpf}{2} \left( \aa e^{i\varphi} - \aad e^{-i\varphi} - \aa e^{-i(2\omega t+\varphi)} + \aad e^{i(2\omega t+\varphi)} \right) \\
&\approx \dfrac{iF_0 \xzpf}{2} \left( \aa e^{i\varphi} - \aad e^{-i\varphi} \right) \fcomma \label{eq:HFprime}
\end{align}
where we used $\hat{U}_0^\dagger \aa \hat{U}_0 = \aa e^{-i\omega t}$ and, in going to the last line, took the rotating-wave approximation (RWA) to neglect the fast-rotating terms oscillating at $2\omega$.
Absorbing a minus sign by re-defining the drive phase as $\varphi=\pi - \phi$ and considering a possibly time-dependent (but slowly varying) force amplitude $F_0(t)$, we obtain Eq.~\eqref{eq:dispH} of the main text
\begin{equation}\label{supp:dispH}
    \hat H/\hbar = i \Omega(t) \left(e^{i\phi}\aad-e^{-i\phi}\aa\right) \;, \qquad\text{with}\qquad 
    \Omega(t) = \dfrac{F_0(t) \xzpf}{2 \hbar} \fperiod
\end{equation}
For a time-dependent drive amplitude $\Omega(t)$ at fixed phase $\phi$, note that $[\hat H(t),\hat H(t')]=0$. The time-evolution operator therefore does not require time ordering and is given by
\begin{align}
\hat U(t) &= \exp\left[-\frac{i}{\hbar}\int_0^t d\tau\,\hat H(\tau)\right] \notag\\
&= \exp\left[\left(\int_0^t d\tau\,\Omega(\tau)\right)\left(e^{i\phi}\aad-e^{-i\phi}\aa\right)\right] \notag\\
&= \exp\left[d(t)\left(e^{i\phi}\aad-e^{-i\phi}\aa\right)\right] \equiv \hat D\left(d(t)\,e^{i\phi}\right) \fcomma \label{eq:dispHFprime}
\end{align}
where $\hat D(d(t)\,e^{i\phi})$ is the displacement operator realizing a phase-space displacement of amplitude
\begin{equation}\label{eq:alpha_F0_tdep}
d(t) \equiv \int_0^t d\tau\,\Omega(\tau)
= \frac{\xzpf}{2\hbar}\int_0^t d\tau\,F_0(\tau) \fperiod
\end{equation}
Note here that the action of the displacement operator $\hat D(\alpha)$ with $\alpha = d e^{i\phi}$ is also immediately connected to the unitary encoding $\rho_d=e^{-id\hat G(\phi)}\rho\,e^{id\hat G(\phi)}$ discussed in the main text, with generator
\begin{equation}\label{supp:generatorG}
    \hat G(\phi) = i \left(e^{i\phi}\aad-e^{-i\phi}\aa\right)  \fperiod
\end{equation}

Assume for simplicity the case of a constant driving force amplitude, $F_0(t)=F_0$, such that Eq.~\eqref{eq:alpha_F0_tdep} reduces to 
\begin{equation}\label{eq:alpha_F0} 
    d(t)=\Omega\, t= \dfrac{F_0\xzpf t}{2\hbar} \;.
\end{equation}
As $F_0$ and $d$ are linearly proportional to each other, the corresponding uncertainties are related by simple error propagation,
\begin{equation}
    \Delta F_0 = \dfrac{2 \hbar}{\xzpf t} \, \Delta d \fperiod
    \label{eq:DeltaF0_from_theta}
\end{equation}
The uncertainty $\Delta d$ achievable using $M$ independent repetitions of the protocol is in turn bounded by the quantum Cramér--Rao bound $\Delta d \geq 1/\sqrt{M\,F_Q[ d]}$,
with $F_Q[ d]$ the single-shot QFI of the probe state employed.
Combining Eq.~\eqref{eq:DeltaF0_from_theta} with the QCRB gives
\begin{equation}
    \Delta F_0 \geq \dfrac{2 \hbar}{\xzpf t} \dfrac{1}{\sqrt{M\, F_Q[d]}} \fperiod
    \label{eq:DeltaF0_QCRB}
\end{equation}
Collecting $M$ measurements requires a total time $T = M\, t_\text{cycle} = M\,(t + t_\text{dead})$, where $t_\text{dead}$ is the extra time needed to, e.g., initialize, read out, and reset the probe system in each measurement shot. Substituting $M=T/t_\text{cycle}$ into Eq.~\eqref{eq:DeltaF0_QCRB} gives, after multiplying both sides by $\sqrt{t_\text{cycle}}$,
\begin{equation}
    \Delta F_0 \, \sqrt{t_\text{cycle}} \;\geq\; \dfrac{2 \hbar}{\xzpf} \dfrac{t_{\text{cycle}}}{t} \dfrac{1}{\sqrt{T\, F_Q[d]}} \fperiod
    \label{eq:sensitivity_general}
\end{equation}
As expected, the actual measurement uncertainty decreases as $\Delta F_0(T)\propto1/\sqrt{T}$ for independent repetitions. The standard time-normalized force sensitivity, usually quoted in units of force$/\sqrt{\mathrm{Hz}}$, is defined as
$S_F\equiv\Delta F_0(T)\sqrt{T}$. Multiplying Eq.~\eqref{eq:sensitivity_general} by $\sqrt{T/t_\text{cycle}}$ gives
\begin{equation}
    S_F \equiv \Delta F_0(T)\sqrt{T} \;\geq\; \dfrac{2\hbar}{\xzpf}\, \dfrac{\sqrt{t_\text{cycle}}}{t}\, \dfrac{1}{\sqrt{F_Q[d]}} \fperiod
    \label{eq:sensitivity_asymptotic}
\end{equation}
For independent repetitions, the right-hand side is independent of the total averaging time $T$. Relative to zero dead time at fixed interrogation time $t$, the sensitivity is degraded by the square root of the inverse duty cycle, $\sqrt{t_\text{cycle}/t}$, so that reducing $t_\text{dead}$ improves the sensitivity.

\subsection{Sensitivity estimates}

To make a concrete estimate, consider the oscillator frequency to be $\omega/2\pi=\omega_a/2\pi=\unit{5.088}{GHz}$, and its effective mass $\mathfrak{m}=\unit{16}{\mu g}$ \cite{bild_schrodinger_2023}, resulting in $\xzpf = \unit{3.2\cdot 10^{-19}}{m}$.
The energy relaxation rate is $\Gamma=1/T_{1,a}$, with $T_{1,a}=\unit{196}{\micro\second}$, thus requiring a significant dead time after each measurement shot for passive reset of the mode.
Taking $t_\text{cycle}=\unit{1}{ms}$ and $t=\unit{6}{\micro\second}$ in Eq.~\eqref{eq:sensitivity_asymptotic} gives for $F_Q=4$ the classical bound $S_F \simeq \unit{1736}{fN/\sqrt{Hz}}$, which corresponds to $\Delta F_0(\unit{9}{days},\ket{0})=\unit{2.0}{fN}$ when averaging over $T=\unit{9}{days}$. The higher Fock states prepared in this work have demonstrated $F=20$, thus providing a $\sqrt{5}\simeq 2.2$ better performance in sensitivity, with $\Delta F_0(\unit{9}{days},\ket{4})=\unit{0.9}{fN}$ over the same averaging time.
Dividing by the mode's effective mass $\mathfrak{m}=\unit{16}{\mu g}$, these bounds correspond to acceleration sensitivities $\Delta a=\Delta F_0/\mathfrak{m}$ of $\Delta a(\unit{9}{days},\ket{0}) = \unit{1.2\cdot10^{-7}}{m/s^2}$ and $\Delta a(\unit{9}{days},\ket{4})=\unit{5.6\cdot10^{-8}}{m/s^2}$, respectively.

To make a comparison, Ref.~\cite{omahen_ultracold_2026} uses a device of the same type to sense a resonant force through the steady-state occupation of the mechanical mode.
For a constant drive in the presence of loss at rate $\Gamma'=1/T_1'$, Eq.~\eqref{suppeq:dOft} gives $d(t\to\infty)=2\Omega/\Gamma'$, and the mode remains coherent.
Using $\Omega=F_0\xzpf/2\hbar$, the steady-state occupation is
\begin{equation}\label{supp:nss}
    \bar{n}_\text{ss}=\left(\frac{2\Omega}{\Gamma'}\right)^{\!2}=\left(\frac{F_0\xzpf}{\hbar\Gamma'}\right)^{\!2} \fperiod
\end{equation}
In that experiment the occupation is inferred from a single qubit measurement per repetition.
The resulting single-shot resolution on $\bar n$ is therefore of order one phonon, $\sigma_n=\mathcal{O}(1)$.
Averaging $M'$ repetitions over a total time $T$ gives $\delta n = \sigma_n/\sqrt{M'} \equiv S_n/\sqrt{T}$, so that both quantities follow from the reported statistics: with $\delta n=5.5\times10^{-5}$ obtained from $M'=3.24\times10^{8}$ repetitions over $T=\unit{9}{days}$ \cite{omahen_ultracold_2026}, corresponding to a cycle time $t'_\text{cycle}=T/M'\simeq\unit{2.4}{ms}$, one finds $\sigma_n\simeq1$ and $S_n=\delta n \sqrt{T}\simeq\unit{4.9\cdot10^{-2}}{Hz^{-1/2}}$.
Requiring $\bar n_\text{ss}\leq\delta n$ gives the smallest force detectable after an averaging time $T$,
\begin{equation}\label{supp:DeltaF0_energy}
    \Delta F_0(T)=\frac{\hbar\Gamma'}{\xzpf}\sqrt{S_n}\;T^{-1/4}
    =\frac{\hbar\Gamma'}{\xzpf}\sqrt{\sigma_n}\left(\frac{t'_\text{cycle}}{T}\right)^{\!1/4} \fperiod
\end{equation}
Note that $\Delta F_0$ decreases only as $T^{-1/4}$, so that no time-normalized sensitivity can be defined for this scheme: the quantity $\Delta F_0\sqrt{T}$, independent of $T$ in Eq.~\eqref{eq:sensitivity_asymptotic}, grows here as $T^{1/4}$.
Two features distinguish this from Eq.~\eqref{eq:sensitivity_asymptotic}.
First, the dead time enters only through the fourth root of $t'_\text{cycle}$, and it costs repetitions but not signal: the mode is continuously in its steady state, so the waiting time between shots does not reduce the accumulated displacement.
This is in contrast with our protocol, where the signal accumulates only during the interrogation window and the duty cycle enters as $\sqrt{t_\text{cycle}}/t$.
Second, the $T^{-1/4}$ scaling is not intrinsic to energy detection.
It arises because the occupation is estimated against a floor $\sigma_n$ that does not shrink with the signal, whereas phonon counting measures the population of $\mu\geq1$ against a vanishing background, keeping the Fisher information finite as $d\rightarrow0$ and restoring $\Delta F_0\propto T^{-1/2}$.
With $1/\Gamma'=T_1'=\unit{112}{\micro\second}$ \cite{omahen_ultracold_2026}, Eq.~\eqref{supp:DeltaF0_energy} gives $\Delta F_0(\unit{1}{s})=\unit{648}{fN}$ when averaging over $T=\unit{1}{s}$, and $\Delta F_0(\unit{9}{days})=\unit{22}{fN}$ for the $T=\unit{9}{days}$ actually measured.
Equivalently, these bounds correspond to acceleration sensitivities of $\Delta a(\unit{1}{s}) = \unit{4.1\cdot10^{-5}}{m/s^2}$ and $\Delta a(\unit{9}{days})=\unit{1.4\cdot10^{-6}}{m/s^2}$, respectively.

In Tab.~\ref{supptab:accel} we compare the sensitivity of our device to other force and acceleration sensors reported in the literature. The comparison spans more than nine orders of magnitude in frequency, with mass and bandwidth differing widely across platforms. In the GHz band, HBAR devices stand essentially alone.

\begingroup
\footnotesize
\setlength{\tabcolsep}{8pt}
\renewcommand{\arraystretch}{1.3}
\begin{table}[t]
\caption{\justifying \textbf{Sensitivities across mechanical sensing platforms, ordered by decreasing frequency.}
$S_a$ is the acceleration noise spectral density, as reported or converted from the quoted force sensitivity via $S_a=S_F/\mathfrak{m}$. Unless stated otherwise, the estimation uncertainty after an averaging time $T$ follows $\Delta F(T)=S_F/\sqrt{T}$ or $\Delta a(T)=S_a/\sqrt{T}$, assuming ideal $1/\sqrt{T}$ averaging. 
For most classical instruments this might not be achievable in practice, since drift and systematics dominate beyond $10^2$--$10^3\,$s.
$^\dagger$Ref.~\cite{omahen_ultracold_2026} measures an observable quadratic in the signal, giving $\Delta a\propto T^{-1/4}$. The quoted $S_a$ is its value at $T=\unit{9}{days}$.
$^\ddagger$Projection obtained by scaling the zero-point fluctuation, $\xzpf=\sqrt{\hbar/2\mathfrak{m}\omega}\propto\omega^{-1/2}$, at fixed effective mass $\mathfrak{m}=\unit{16}{\mu g}$, interrogation time $t=\unit{6}{\micro\second}$, unit duty cycle, and Fisher information $F=20$. Since $S_a\propto1/\xzpf$, the sensitivity improves as $\omega^{-1/2}$.}
\label{supptab:accel}
\begin{tabular}{llllll}
\hline\hline
\makecell[l]{Oscillator / platform} & Mass & Frequency &
\makecell[l]{$S_F$\\(N\,Hz$^{-1/2}$)} &
\makecell[l]{$S_a$\\(m\,s$^{-2}$\,Hz$^{-1/2}$)} & Ref. \\
\hline
HBAR phonon mode, Fock $\ket{4}$ probe        & $\unit{16}{\mu g}$   & $\unit{5.09}{GHz}$  & $7.7\times10^{-13}$ & $4.8\times10^{-5}$ & this work \\
HBAR phonon mode, vacuum probe                & $\unit{16}{\mu g}$   & $\unit{5.09}{GHz}$  & $1.7\times10^{-12}$ & $1.1\times10^{-4}$ & this work \\
HBAR, steady-state occupation                 & $\unit{16}{\mu g}$   & $\unit{5.05}{GHz}$  & $1.9\times10^{-11}$ & $1.2\times10^{-3}$\,$^\dagger$ & \cite{omahen_ultracold_2026} \\
Carbon-nanotube resonator ($\unit{1.2}{K}$)   & $\unit{10}{ag}$      & $\unit{4.2}{MHz}$   & $1.2\times10^{-20}$ & $1.2$ & \cite{moser_ultrasensitive_2013} \\
Si$_3$N$_4$ nanobeam, near-field optomech.    & $\unit{9}{pg}$       & $\unit{2.88}{MHz}$  & $7.4\times10^{-17}$ & $8.2\times10^{-3}$ & \cite{gavartin_hybrid_2012} \\
Trapped-ion crystal ($n{=}150$, $^{9}$Be$^{+}$) & $\unit{2.2}{zg}$   & $\unit{1.6}{MHz}$   & $5.8\times10^{-24}$ & $2.6$ & \cite{gilmore_quantum-enhanced_2021} \\
Single trapped ion ($^{88}$Sr$^{+}$)          & $\unit{146}{yg}$     & $\unit{1.13}{MHz}$  & $2.8\times10^{-20}$ & $1.9\times10^{5}$ & \cite{shaniv_quantum_2017} \\
Trapped-ion crystal ($n{=}60$, $^{9}$Be$^{+}$) & $\unit{0.90}{zg}$   & $\unit{867}{kHz}$   & $3.9\times10^{-22}$ & $4.3\times10^{2}$ & \cite{biercuk_ultrasensitive_2010} \\
Si$_3$N$_4$ membrane optomechanical accel.    & $\unit{12}{n g}$ & $\unit{39}{kHz}$ & $1.2\times10^{-18}$ & $9.8\times10^{-8}$ & \cite{manley_optomechanical_2025} \\
Photonic-crystal optomechanical accel.        & $\sim\unit{10}{ng}$  & $\unit{28}{kHz}$    & $9.8\times10^{-16}$ & $9.8\times10^{-5}$ & \cite{krause_high-resolution_2012} \\
Fused-silica optomechanical reference accel.  & $\unit{25}{mg}$  & $\unit{10}{kHz}$ & $2.5\times10^{-11}$ & $9.8\times10^{-7}$ & \cite{guzman_high_2014} \\
Fabry--P\'erot optomechanical accel.          & $\unit{11.1}{mg}$    & $\unit{10}{kHz}$ & $3.5\times10^{-12}$ & $3.1\times10^{-7}$ & \cite{zhou_broadband_2021} \\
Commercial servo accelerometer (Si-Flex)      & $\sim\unit{1}{g}$    & DC--$\unit{1}{kHz}$ & $2.9\times10^{-9}$ & $2.9\times10^{-6}$ & \cite{laine_high-sensitivity_2014} \\
Advanced LIGO test mass                       & $\unit{40}{kg}$      & $\unit{100}{Hz}$    & $6.4\times10^{-13}$ & $1.6\times10^{-14}$ & \cite{aasi_advanced_2015} \\
Optically levitated silica sphere             & $\sim\unit{1}{ng}$   & $\unit{50}{Hz}$ & $9.5\times10^{-19}$ & $9.3\times10^{-7}$ & \cite{monteiro_force_2020} \\
MEMS gravimeter (quasi-zero stiffness)        & $\sim\unit{10}{mg}$  & $\unit{3.1}{Hz}$     & $8.0\times10^{-13}$ & $8.0\times10^{-8}$ & \cite{tang_high-sensitivity_2019} \\
MEMS gravimeter (Wee-g)                       & $\sim\unit{10}{mg}$  & $\unit{2.3}{Hz}$    & $4.0\times10^{-12}$ & $4.0\times10^{-7}$ & \cite{middlemiss_measurement_2016} \\
MEMS gravimeter (force-balanced)              & $\sim\unit{10}{mg}$  & $\unit{0.6}{Hz}$    & $1.0\times10^{-14}$ & $1.0\times10^{-9}$ & \cite{li_force_2026} \\
LISA Pathfinder test masses                   & $\unit{1.93}{kg}$    & $\unit{2}{mHz}$     & $3.4\times10^{-15}$ & $1.7\times10^{-15}$ & \cite{armano_beyond_2018} \\
Superconducting gravimeter (GWR iGrav)        & $\sim\unit{5}{g}$    & $\unit{1}{mHz}$     & $1.5\times10^{-11}$ & $3.0\times10^{-9}$ & \cite{gwr_igrav} \\
Cold-atom gravimeter (laboratory)             & $\unit{133}{u}$       & free fall           & $4.9\times10^{-32}$ & $2.2\times10^{-7}$ & \cite{peters_high-precision_2001} \\
Cold-atom gravimeter (compact, field)         & $\unit{87}{u}$       & free fall           & $3.9\times10^{-31}$ & $2.7\times10^{-6}$ & \cite{bodart_cold_2010} \\
\hline
HBAR mode at $\unit{100}{MHz}$ & $\unit{16}{\mu g}$   & $\unit{100}{MHz}$   & $8.4\times10^{-15}$ & $5.3\times10^{-7}$ & projected$^\ddagger$ \\
HBAR mode at $\unit{10}{MHz}$ & $\unit{16}{\mu g}$   & $\unit{10}{MHz}$    & $2.7\times10^{-15}$ & $1.7\times10^{-7}$ & projected$^\ddagger$ \\
HBAR mode at $\unit{1}{MHz}$ & $\unit{16}{\mu g}$   & $\unit{1}{MHz}$     & $8.4\times10^{-16}$ & $5.3\times10^{-8}$ & projected$^\ddagger$ \\
\hline\hline
\end{tabular}
\end{table}
\endgroup

\clearpage
\newpage

\section{Quantum metrology for unitary evolutions}
\label{sec:theory_lossless}

In this section, we briefly review the key concepts of quantum metrology used in the main text, with particular emphasis on the special case of unitary evolutions. Non-unitary evolutions relevant to our experiment are discussed in the following section.

\subsection{Classical and quantum Cramér--Rao bounds}

In a parameter estimation task, an unknown parameter $d$ is encoded into the state of a probe system, $\rho\rightarrow\rho_d$, and information about $d$ is subsequently extracted by performing a measurement described by a positive operator-valued measure (POVM) $\{\hat E_\mu\}$, with outcomes $\mu$ occurring with probability $P(\mu|d)=\Tr[\hat E_\mu \rho_d]$.
From a set of $M$ independent and identically distributed outcomes $\bm{\mu} = \{\mu_i\}_{i=1}^M$, an estimator $\Theta(\bm{\mu})$ is constructed to infer $d$.
For any unbiased estimator, the achievable precision is bounded by the classical Cramér--Rao bound (CRB) \cite{fadel_quantum_2025}
\begin{equation}\label{suppeq:CCRB}
    \var{\Theta} \geq \dfrac{1}{M F[d]} \fcomma
\end{equation}
where $F[d]$ is the classical Fisher information (FI) associated with the chosen measurement,
\begin{equation}\label{supp:CFI}
    F[d] = \sum_\mu \dfrac{1}{P(\mu|d)} \left(\dfrac{\partial P(\mu|d)}{\partial d}\right)^{2} \fperiod
\end{equation}
Since $F[d]$ depends on the measurement, one can further ask what is the ultimate precision allowed by quantum mechanics, optimized over all possible POVMs.
This is quantified by the quantum Fisher information (QFI) $F_Q[d]$, obtained by maximizing $F[d]$ over $\{\hat E_\mu\}$, and formally defined through the symmetric logarithmic derivative (SLD) operator $\hat L_d$, implicitly given by
\begin{equation}
    \partial_d \rho_d = \dfrac{1}{2}\left(\hat L_d \rho_d + \rho_d \hat L_d\right) \fcomma
\end{equation}
as $F_Q[d]=\Tr[\rho_d \hat L_d^2]$.
Since $F[d]\leq F_Q[d]$ for any measurement, Eq.~\eqref{suppeq:CCRB} implies the quantum Cramér--Rao bound (QCRB)
\begin{equation}\label{eq:QCRB}
    \var{\Theta} \geq \dfrac{1}{M F[d]} \geq \dfrac{1}{M F_Q[d]} \fperiod
\end{equation}
Writing $\rho_d=\sum_n \lambda_n \ket{\lambda_n}\bra{\lambda_n}$ in its eigenbasis, the QFI takes the explicit form
\begin{equation}
F_Q[d] = 2\sum_{\substack{\ell,n \\ \lambda_\ell+\lambda_n>0}}
\dfrac{\left|\langle \lambda_\ell|\partial_d\rho_d|\lambda_n\rangle\right|^2}
{\lambda_\ell+\lambda_n} \fcomma
\end{equation}
which for the common case of unitary parameter encoding, $\rho_d = e^{-id \hat G}\rho\,e^{id \hat G}$ with Hermitian generator $\hat G$, simplifies to
\begin{equation}\label{eq:FQunitary}
F_Q[d] = 2\sum_{\substack{\ell,n \\ \lambda_\ell+\lambda_n>0}}
\dfrac{(\lambda_\ell-\lambda_n)^2}{\lambda_\ell+\lambda_n}\,
|\langle \lambda_\ell|\hat G|\lambda_n\rangle|^2 \fperiod
\end{equation}
For a pure probe state $\rho=\ket{\psi}\bra{\psi}$, Eq.~\eqref{eq:FQunitary} reduces to the familiar variance formula
\begin{equation}\label{eq:FQpure}
    F_Q[d] = 4\,\var{\hat G}_{\ket{\psi}} = 4\left(\avg{\hat G^2}_{\ket{\psi}} - \avg{\hat G}_{\ket{\psi}}^2\right) \fcomma
\end{equation}
independent of $d$, since the encoding unitary is generated by $\hat G$ itself and therefore leaves all moments of $\hat G$ invariant. 
A projective measurement in the eigenbasis of $\hat L_d$ provides a locally optimal measurement attaining the QFI, although the optimal measurement need not be unique, and, for pure states undergoing unitary encoding, in the asymptotic limit of many repetitions this saturation can always be achieved.

\subsection{Phase-insensitive displacement sensing and optimality of Fock states}

We now specialize the previous quantum sensing framework to the estimation of a phase-space displacement amplitude with a bosonic mode.
As derived in detail in Sec.~\ref{supp:resDrive}, resonantly driving an oscillator for a time $t$ implements the displacement operator $\hat D(d e^{i\phi})$, with $d=\int d\tau\,\Omega(\tau)$ in the ideal, lossless limit (see Eq.~\eqref{eq:alpha_F0}).
This can equivalently be viewed as a unitary encoding of the parameter $d$ through the generator $\hat G(\phi)$ of Eq.~\eqref{supp:generatorG}, for a fixed and known phase-space direction $\phi$.
For the scenario we study, we are further interested in considering this direction to be unknown, or fluctuating from shot-to-shot, and optimize the state in order to maximize the sensitivity in estimating $d$.

For this phase-insensitive displacement sensing task, the worst-case QFI of a state $\rho$ with mean phonon number $\bar{n}=\avg{\aad\aa}$ obeys \cite{fadel_quantum_2025}
\begin{equation}\label{supp:avQFI}
    \min_{\phi} F_Q[d,\{\rho,\hat G(\phi)\}] \leq 4(2\bar{n}+1) \fperiod
\end{equation}
Fock states $\ket{n}$ saturate this bound for any $n$ and $\phi$: using $\avg{\aa}_n=\avg{\aa^2}_n=\avg{\aad{}^2}_n=0$ and $\avg{\aad\aa}_n=n$, one finds $\avg{\hat G(\phi)}_n=0$ and $\avg{\hat G(\phi)^2}_n = - e^{2i\phi}\avg{(\aad)^2}_n - e^{-2i\phi}\avg{\aa^2}_n + \avg{\aa\aad+\aad\aa}_n = 2n+1$. 
Using the QFI expression for pure states Eq.~\eqref{eq:FQpure}, we thus have
\begin{equation}\label{eq:QFI_fock_main}
    F_Q[d,\{\ket{n},\hat G(\phi)\}] = 4(2n+1) \fcomma
\end{equation}
independent of both the displacement magnitude $d$ and its phase $\phi$. 
The latter is a consequence of the rotational symmetry of the Fock-state Wigner function in phase space, unlike, e.g., squeezed or coherent states \cite{fadel_quantum_2025}.
Fock states are therefore optimal probes for phase-insensitive displacement sensing at fixed energy, saturating Eq.~\eqref{supp:avQFI} for any $n$. We further show in Sec.~\ref{sec:optmeas_lossless} that number-resolved (Fock-basis) detection is an optimal measurement for displaced Fock states, i.e.\ it saturates $F[d]=F_Q[d]=4(2n+1)$, which motivates our choice of resonant phonon-number (RPN) readout.
This scaling of the QFI with the phonon number $n$ underlies the quantum-enhanced sensitivity reported in this work, since increasing $n$ increases the achievable precision beyond what is possible with the vacuum ($F_Q=4$) or with any other classical state.

\subsection{Optimality of Fock-population measurements}
\label{sec:optmeas_lossless}

We now want to show that for a displaced Fock state, a measurement in the Fock basis is optimal for estimating the displacement amplitude, meaning that it results in a classical FI that coincides with the QFI Eq.~\eqref{eq:QFI_fock_main}.
To this end, let us introduce the displaced Fock states
\begin{equation}
\ket{\psi_\alpha^k} = \hat{D}(\alpha)\ket{k} = \sum_n \psi_n^k(\alpha) \ket{n} \;, \qquad\text{with}\quad \psi_n^k(\alpha) = \langle n|\hat D(\alpha)|k\rangle \;,
\end{equation}
and the measurement probability of detecting $\ell$ excitations as
\begin{equation}
p_\ell^k(\alpha)=\vert\braket{\ell|\psi_\alpha^{k}}\vert^2 = |\langle \ell|\hat D(\alpha)|k\rangle|^2 = e^{-|\alpha|^2} \frac{\min(k,\ell)!}{\max(k,\ell)!} |\alpha|^{2|k-\ell|} \left[ L^{(|k-\ell|)}_{\min(k,\ell)}(|\alpha|^2) \right]^2 \;.
\end{equation}
Note this expression depends only on the modulus of $\alpha$, and not on its phase. In fact, identifying $\alpha = d e^{i\phi}$, we have $\hat D(d e^{i\phi})=e^{-id\hat G(\phi)}=\hat R^\dagger(\phi)\hat D(d) \hat R(\phi)$, with $\hat G(\phi) = i \left(e^{i\phi}\aad-e^{-i\phi}\aa\right)$ the generator introduced in Eq.~\eqref{supp:generatorG} and $\hat R(\phi)=e^{-i\phi \aad \aa}$ the phase-space rotation operator, as follows directly from $\hat R^\dagger(\phi) \aad \hat R(\phi)= \aad e^{i\phi}$ and $\hat R^\dagger(\phi)\aa \hat R(\phi)=\aa e^{-i\phi}$. Since $\hat R(\phi)\ket{k}=e^{-ik\phi}\ket{k}$, the Fock-basis amplitudes obey
\begin{equation}\label{eq:Dmtxelements}
\braket{\ell|\hat D(d e^{i\phi})|k} = e^{i(\ell-k)\phi}\braket{\ell|\hat D(d)|k} = e^{i(\ell-k)\phi}c_{\ell k}(d) \;,
\end{equation}
where $c_{\ell k}(d)=\langle \ell|\hat{D}(d)|k\rangle\in\mathbb{R}$, because for real $d$ the matrix $d(\aad-\aa)$ appearing in the exponent is real and skew-symmetric in the Fock basis, so $\hat{D}(d)=\exp[d(\aad-\aa)]$ is a real orthogonal (i.e., real-valued unitary) matrix. We may therefore write the phonon-counting probabilities and their derivative, for the fixed input label $k$ (henceforth $c_\ell(d)\equiv c_{\ell k}(d)$), as
\begin{equation}
p_\ell(d)=c_\ell^2(d), \qquad \partial_d p_\ell(d)=2c_\ell(d)c_\ell'(d) \;.
\end{equation}
The classical Fisher information associated with number-resolved detection is therefore
\begin{equation}\label{eq:FockCFI}
F_C[\ket{k}] = \sum_\ell \frac{[\partial_d p_\ell(d)]^2}{p_\ell(d)} = 4\sum_\ell [c_\ell'(d)]^2 \fperiod
\end{equation}
At isolated values of $d$ for which $p_\ell(d)=0$, the corresponding Fisher-information term is understood by continuity, taking the limit $d\to d_0$ before evaluating the ratio. To evaluate the QFI, we regard the displaced state as a one-parameter family along a fixed phase-space direction $\phi$, defining $\ket{\psi_{d,\phi}^k}\equiv\ket{\psi_{\alpha=d e^{i\phi}}^k}=\hat D(d e^{i\phi})\ket{k}$. The pure-state QFI for estimating $d$ is then
\begin{equation}
F_Q[d;\ket{k}] = 4\left[\braket{\partial_d\psi_{d,\phi}^k|\partial_d\psi_{d,\phi}^k} - \left|\braket{\psi_{d,\phi}^k|\partial_d\psi_{d,\phi}^k}\right|^2\right] \fperiod
\end{equation}
Using Eq.~\eqref{eq:Dmtxelements}, the derivative at fixed $\phi$ is
\begin{equation}
\partial_d\ket{\psi_{d,\phi}^k}=\sum_\ell e^{i(\ell-k)\phi}c_\ell'(d)\ket{\ell} \fcomma
\end{equation}
and therefore
\begin{equation}
\braket{\partial_d\psi_{d,\phi}^k|\partial_d\psi_{d,\phi}^k}=\sum_\ell[c_\ell'(d)]^2 \fperiod
\end{equation}
For the second term, normalization $\sum_\ell c_\ell^2(d)=1$ gives
\begin{equation}
\braket{\psi_{d,\phi}^k|\partial_d\psi_{d,\phi}^k}=\sum_\ell c_\ell(d)c_\ell'(d)=\frac{1}{2}\partial_d\sum_\ell c_\ell^2(d)=0 \fperiod
\end{equation}
Hence $F_Q[d;\ket{k}]=4\sum_\ell[c_\ell'(d)]^2$, which coincides with the classical FI in Eq.~\eqref{eq:FockCFI}. Thus, number-resolved detection is an optimal projective measurement for estimating the displacement amplitude $d$ along any fixed phase-space direction $\phi$. Rotational covariance ensures that changing $\phi$ only multiplies each Fock amplitude by the $d$-independent phase factor $e^{i(\ell-k)\phi}$, leaving the phonon-number probabilities, and hence the Fisher information for $d$, unchanged.

This result applies to the displacement amplitude estimation. If instead the unknown parameter is the displacement angle $\phi$, the probabilities $p_\ell(d)$ are independent of $\phi$, and Fock-basis measurements do not in general saturate the corresponding QFI.

\bigskip
As Fock population measurements are optimal for sensing displacements using a pure Fock state, we might ask whether they are also optimal when the probe state is an arbitrary Fock-diagonal state. To show that, in general, this is not the case, consider $\rho_0=\sum_n \lambda_n |n\rangle\langle n|$, with $\lambda_n\geq0$ and $\sum_n \lambda_n=1$.
Defining $\rho_{d} = \hat{D}(de^{i\phi}) \rho_0\hat{D}^\dagger(de^{i\phi})$, we find using Eq.~\eqref{eq:Dmtxelements} that the Fock-basis probabilities are
\begin{equation}
p_\ell(d) = \langle \ell| \rho_{d}|\ell\rangle = \sum_n \lambda_n c_{\ell n}^2(d) \fcomma
\end{equation}
and the classical FI is therefore
\begin{equation}\label{eq:FCdiag}
F_C = \sum_\ell \frac{[\partial_d p_\ell(d)]^2}{p_\ell(d)} = 4\sum_\ell \frac{\left[\sum_n \lambda_n c_{\ell n}(d)c'_{\ell n}(d)\right]^2}{\sum_n \lambda_n c_{\ell n}^2(d)} \fperiod
\end{equation}
This expression is not generally equal to the mixed-state QFI. To see this, note that the generator $\hat G(\phi)$ Eq.~\eqref{supp:generatorG} couples only neighboring Fock levels:
\begin{equation}
\langle j|\hat G(\phi)|l\rangle = i e^{i\phi}\sqrt{l+1}\,\delta_{j,l+1} - i e^{-i\phi}\sqrt{l}\,\delta_{j,l-1}
\ \Longrightarrow
|\langle j|\hat G(\phi)|l\rangle|^2 = (l+1) \,\delta_{j,l+1} + l \,\delta_{j,l-1} \fperiod
\end{equation}
For $\rho_{d}$ generated by this unitary encoding of $d$ at fixed, known $\phi$, the exact QFI is given by the mixed-state expression
\begin{equation}
F_Q = 2\sum_{\substack{j,l \\ \lambda_j+\lambda_l>0}} \frac{(\lambda_j - \lambda_l)^2}{\lambda_j + \lambda_l} \, |\langle j| \hat G(\phi) |l\rangle|^2 \fperiod
\end{equation}
Since only pairs $l=j\pm1$ contribute, and the summand is symmetric under $j\leftrightarrow l$ [both $(\lambda_j-\lambda_l)^2$ and $|\langle j|\hat G(\phi)|l\rangle|^2$ are], each unordered pair $\{n,n+1\}$ contributes twice to the sum over ordered pairs $(j,l)$, giving
\begin{equation}\label{eq:FQdiag}
F_Q = 4\sum_{\substack{n=0 \\ \lambda_{n+1}+\lambda_n>0}}^{\infty} (n+1)\, \frac{(\lambda_{n+1}-\lambda_n)^2}{\lambda_{n+1}+\lambda_n} \fperiod
\end{equation}

The inequivalence between Eq.~\eqref{eq:FCdiag} and Eq.~\eqref{eq:FQdiag} is particularly transparent in the limit $d\to0^+$. For outcomes with $\lambda_\ell>0$, the population derivatives vanish linearly in $d$ and their contribution to the classical FI vanishes. Outcomes with $\lambda_\ell=0$, however, require taking the limit of $[\partial_d p_\ell(d)]^2/p_\ell(d)$ before setting $d=0$, and can give a finite contribution. Hence $F_C(0^+)$ is not generically zero for Fock-diagonal states with non-full support, reflecting the fact that the population change induced by an infinitesimal displacement is $\mathcal O(d^2)$.
By contrast, $F_Q$ in Eq.~\eqref{eq:FQdiag} is manifestly independent of $d$, as expected for a unitary encoding generated by the parameter-independent operator $\hat G(\phi)$, and is generally nonzero already at $d=0$ whenever $\lambda_{n+1}\neq \lambda_n$ for at least one $n$.

Thus number-resolved detection does not, in general, saturate the QFI for displacement amplitude estimation of a Fock-diagonal mixed state. The equality $F_C=F_Q$ holds identically for pure Fock states $\lambda_n=\delta_{n,k}$ (Sec.~\ref{sec:optmeas_lossless} above), but is not generic for mixed Fock-diagonal states.

\clearpage
\newpage

\section{Displacement sensing with losses}

The results presented in Sec.~\ref{sec:theory_lossless} assume an ideal, noiseless unitary evolution. In a realistic scenario the probe system is inevitably coupled to the environment through undesired channels, resulting in decay and decoherence of the probe state and thus reducing the sensitivity. 
In fact, the loss-only evolution maps an initial Fock state to a Fock-diagonal mixed state, $\rho=\ket{k}\bra{k}\rightarrow\rho(t)=\sum_n \lambda_n(t)\ket{n}\bra{n}$, so that the QFI must now be evaluated using the general mixed-state expression Eq.~\eqref{eq:FQunitary} rather than the pure-state formula Eq.~\eqref{eq:FQpure}: the two generally disagree once $\rho_d$ departs from a pure state.

\subsection{Quantum channel description of a displacement with losses}

Consider the dynamics generated by the time-dependent Hamiltonian $\hat H/\hbar = i \Omega(t) \left(e^{i\phi}\aad-e^{-i\phi}\aa\right)$ in the presence of losses described by coupling the mode to a zero-temperature bath at rate $\Gamma=1/T_1$, where $T_1$ is the energy relaxation time of the mode. The evolution is governed by the Lindblad master equation
\begin{equation}
  \dot\rho=-\frac{i}{\hbar}[\hat H(t),\rho]+\Gamma\left(\hat a\rho\hat a^\dagger-\frac{1}{2}\{\hat a^\dagger\hat a,\rho\}\right) \;,
  \label{eq:master}
\end{equation}
which for an evolution time $t$ defines a quantum channel (completely positive trace-preserving map) $\mathcal{E}_t[\rho]$. For arbitrary $\Omega(t)$, this channel can be written exactly as a pure-loss channel followed by a displacement,
\begin{equation}\label{supp:chDec}
   \mathcal{E}_t=\mathcal D_{\alpha(t)}\circ\mathcal L_{\eta(t)}
   =\mathcal L_{\eta(t)}\circ\mathcal D_{\alpha(t)/\sqrt{\eta(t)}} \;,
\end{equation}
where
\begin{equation}\label{supp:etabeta}
    \eta(t)=e^{-\Gamma t} \;, \qquad\qquad \alpha(t)= d(t)\,e^{i\phi} \;, \qquad\qquad d(t)=\int_0^t d\tau \,e^{-\Gamma(t-\tau)/2}\Omega(\tau) \;,
\end{equation}
are the transmissivity, effective displacement, and its amplitude, respectively.
Thus, the contribution applied at time $\tau$ is attenuated during the remaining evolution time $(t-\tau)$. In the lossless limit $\Gamma t\rightarrow0$, Eq.~\eqref{supp:etabeta} reduces to $d(t)=\int_0^t d\tau\,\Omega(\tau)$, recovering the unitary displacement generated by the time-dependent drive Eq.~\eqref{eq:alpha_F0_tdep}. For the constant-drive case $\Omega(t)=\Omega$, the integral evaluates to
\begin{equation}\label{suppeq:dOft}
    d(t)=\frac{2\Omega}{\Gamma}\left(1-e^{-\Gamma t/2}\right) \;,
\end{equation}
recovering the result used in the main text.
For $\Gamma t \rightarrow 0$ we have $d(t)\rightarrow  \Omega t$, recovering the ideal unitary displacement of Eq.~\eqref{eq:dispHFprime}.
In the remainder of this section we prove the channel decomposition above, deferring the explicit Kraus-operator representations of $\mathcal L_\eta$ and $\mathcal D_\alpha$ to the two subsections below.

\vspace{2mm}
\textbf{Proof of Eq.~\eqref{supp:chDec}.}
The master equation~\eqref{eq:master} is Gaussian (linear Hamiltonian, amplitude-damping dissipator), and therefore generates a \emph{Gaussian} channel. A single-mode Gaussian channel is uniquely specified by the affine transformation of its first moments and the corresponding linear transformation and state-independent noise addition to its covariance matrix. It is therefore sufficient to verify that $\mathcal E_t$ and the channel composition $\mathcal D_{\alpha(t)}\circ\mathcal L_{\eta(t)}$ have identical Gaussian-channel parameters.

\vspace{2mm}
\emph{First moment.}
In the Heisenberg picture, the quantum Langevin equation associated with Eq.~\eqref{eq:master} reads
\begin{equation}
\frac{d\hat a(t)}{dt} = \Omega(t) e^{i\phi} - \frac{\Gamma}{2}\hat a(t) + \sqrt{\Gamma} \,\hat a_{\rm in}(t) \;,
\label{eq:langevin}
\end{equation}
where $\hat a_{\rm in}(t)$ is a vacuum input-noise operator satisfying $[\hat a_{\rm in}(t),\hat a_{\rm in}^\dagger(t')]=\delta(t-t')$, $\avg{\hat a_{\rm in}(t)\hat a_{\rm in}^\dagger(t')}=\delta(t-t')$, $\avg{\hat a_{\rm in}^\dagger(t)\hat a_{\rm in}(t')}=0$, and $\avg{\hat a_{\rm in}(t)\hat a_{\rm in}(t')}=0$. Equation~\eqref{eq:langevin} is linear and inhomogeneous, and its solution is
\begin{equation}
\hat a(t) = e^{-\Gamma t/2}\hat a(0) + \int_0^t dt'\, e^{-\Gamma(t-t')/2}\Omega(t')e^{i\phi} + \sqrt{\Gamma}\int_0^t dt'\, e^{-\Gamma(t-t')/2}\hat a_{\rm in}(t') \;.
\label{eq:langevin_sol}
\end{equation}
The deterministic drive term is precisely
\begin{equation}
\int_0^t dt'\, e^{-\Gamma(t-t')/2}\Omega(t')e^{i\phi} = d(t) e^{i\phi} =\alpha(t) \;,
\end{equation}
as defined in Eq.~\eqref{supp:etabeta}, while $e^{-\Gamma t/2}=\sqrt{\eta(t)}$. Taking the expectation value of Eq.~\eqref{eq:langevin_sol} (the noise term has zero mean) gives
\begin{equation}
\avg{\hat a}(t) = \sqrt{\eta(t)}\, \avg{\hat a}(0) + \alpha(t) \;,
\label{eq:a_t_affine}
\end{equation}
which is precisely the first-moment map of a pure-loss channel of transmissivity $\eta(t)$ (rescaling $\avg{\aa}(0)\to\sqrt\eta\avg{\aa}(0)$) followed by a unitary displacement by $\alpha(t)$ (shifting $\avg \aa\to\avg \aa+\alpha(t)$), i.e., the affine map generated by $\mathcal D_{\alpha(t)}\circ\mathcal L_{\eta(t)}$.

\vspace{2mm}
\emph{Second moments.}
Define the noise operator $\hat \xi(t) \equiv \sqrt{\Gamma}\int_0^t dt'\, e^{-\Gamma(t-t')/2}\hat a_{\rm in}(t')$, so that $\hat a(t)-\avg{\hat a(t)} = \sqrt{\eta(t)}\left[\hat a(0)-\avg{\hat a(0)}\right] + \hat \xi(t)$. Using the input-noise correlators above,
\begin{equation}
\avg{\hat \xi(t)\hat \xi^\dagger(t)} = \Gamma\int_0^t dt'\, e^{-\Gamma(t-t')} = 1-\eta(t) \;, \qquad
\avg{\hat \xi^\dagger(t)\hat \xi(t)} = \avg{\hat \xi(t)\hat \xi(t)} = 0 \;,
\end{equation}
and $\hat \xi(t)$ is uncorrelated with $\hat a(0)$. This is precisely the noise added by a pure-loss channel $\mathcal L_{\eta(t)}$ acting on a vacuum ancilla: it rescales the input covariance matrix by $\eta(t)$ and adds isotropic vacuum noise $(1-\eta(t))/2$ in each quadrature, with no anomalous (squeezing-like) contribution. Importantly, the time-dependent drive $\Omega(t)$ enters only through the deterministic displacement $\alpha(t)$ and therefore does not affect the covariance matrix. The subsequent unitary displacement $\mathcal D_{\alpha(t)}$ likewise leaves the covariance matrix untouched. Hence the covariance-matrix map induced by $\mathcal E_t$ coincides with that of $\mathcal D_{\alpha(t)}\circ\mathcal L_{\eta(t)}$.
Since $\mathcal E_t$ and $\mathcal D_{\alpha(t)}\circ\mathcal L_{\eta(t)}$ are Gaussian channels with identical first-moment and covariance-matrix transformations, they coincide as channels, proving the first equality in Eq.~\eqref{supp:chDec}.

\vspace{2mm}
\emph{Equivalence of the two orderings.}
The second equality in Eq.~\eqref{supp:chDec} follows by noting that the loss channel is covariant under displacements. Therefore, applying $\mathcal D_{\alpha(t)/\sqrt{\eta(t)}}$ before the loss channel $\mathcal L_{\eta(t)}$ produces, at the level of first moments, $\avg{\aa}(0)\to \avg{\aa}(0)+\alpha(t)/\sqrt{\eta(t)} \to \sqrt{\eta(t)}\avg{\aa}(0) + \alpha(t)$, identical to Eq.~\eqref{eq:a_t_affine}. The covariance-matrix map is unaffected by the ordering, since displacements leave all centered second moments unchanged. Thus
\begin{equation}
\mathcal D_{\alpha(t)}\circ\mathcal L_{\eta(t)} = \mathcal L_{\eta(t)} \circ \mathcal D_{\alpha(t)/\sqrt{\eta(t)}} \;,
\end{equation}
completing the proof. $\blacksquare$

\vspace{2mm}
The explicit Kraus-operator (Fock-basis) representations of the pure-loss channel $\mathcal L_\eta$ and the unitary displacement channel $\mathcal D_\alpha$ used to evaluate Eq.~\eqref{supp:chDec} on arbitrary, in particular Fock-diagonal, input states are derived in the two following subsections, and are subsequently combined in Sec.~\ref{sec:loss_disp_combination} to obtain the phonon-number distribution and the QFI relevant to our sensing protocol.

\subsection{Action of the loss channel}

The pure-loss (amplitude-damping) channel describing coupling of a single bosonic mode to a zero-temperature bath with transmissivity $\eta=e^{-\Gamma t}$ admits the Kraus decomposition
\begin{equation}
\mathcal L_\eta[\rho]
= \sum_{\ell=0}^\infty \hat{K}_\ell \rho \hat{K}_\ell^\dagger , \qquad\text{with}\quad \hat{K}_\ell
= \sqrt{\frac{(1-\eta)^\ell}{\ell!}}\,
 \eta^{\frac{1}{2}\aad \aa} \, \aa^\ell \;.
\end{equation}
Acting on a Fock state $|k\rangle$, one finds
\begin{equation}
\hat{K}_\ell |k\rangle
= \sqrt{\binom{k}{\ell}}\,
(1-\eta)^{\ell/2}\eta^{(k-\ell)/2}\,
|k-\ell\rangle ,
\qquad \ell \le k,
\end{equation}
and zero otherwise. The corresponding matrix elements are therefore
\begin{equation}
\langle n|\hat{K}_\ell|k\rangle
= \sqrt{\binom{k}{\ell}}\,
(1-\eta)^{\ell/2}\eta^{(k-\ell)/2}\,
\delta_{n,k-\ell}.
\end{equation}

Let the initial density operator be $\rho(0)=\sum_{j,k=0}^\infty \rho_{jk}(0)\,|j\rangle\langle k|$.
After the loss channel, $\rho(t)=\mathcal L_\eta[\rho(0)]$, the Fock-basis matrix elements are
\begin{equation}
\rho_{mn}(t)
= \sum_{\ell=0}^\infty
\sqrt{\binom{m+\ell}{\ell}\binom{n+\ell}{\ell}}\,
(1-\eta)^\ell\,\eta^{(m+n)/2}\,
\rho_{m+\ell,n+\ell}(0).
\label{eq:loss_fock_elements}
\end{equation}
For numerical calculations, the sum effectively truncates at the chosen Hilbert-space cutoff.

Note from Eq.~\eqref{eq:loss_fock_elements} that the diagonal elements obey
\begin{equation}
\rho_{nn}(t)
= \sum_{\ell=0}^\infty
\binom{n+\ell}{\ell}(1-\eta)^\ell \eta^{n}\,
\rho_{n+\ell,n+\ell}(0),
\end{equation}
showing explicitly that the output populations depend only on the initial Fock-state populations and not on the coherences.
This result admits a direct interpretation: each of the $k=n+\ell$ initial phonons is transmitted independently with probability $\eta$, so that the output phonon-number distribution $\{\rho_{nn}(t)\}$ is obtained from the initial distribution $\{\rho_{nn}(0)\}$ by binomial thinning.

For an initial Fock state $|k\rangle$ we find that it evolves into the binomial distribution
\begin{equation}
    \rho_{nn}(t) = \binom{k}{n} \eta^n(1-\eta)^{k-n} \;,\qquad \text{for $n\le k$.}
\end{equation}

\subsection{Action of the displacement channel}

The unitary displacement channel with complex amplitude $\alpha = d e^{i\phi}$ acts as
\begin{equation}
\mathcal D_\alpha[\rho]=\hat{D}(\alpha)\rho \hat{D}^\dagger(\alpha),
\qquad
\hat{D}(\alpha)=\exp\left(\alpha \aad - \alpha^* \aa \right).
\end{equation}
Using the Baker--Campbell--Hausdorff identity, the displacement operator can be written in normally ordered form as
\begin{equation}
\hat{D}(\alpha)
= e^{-|\alpha|^2/2}\, e^{\alpha \aad} e^{-\alpha^* \aa}.
\end{equation}
The Fock-basis matrix elements $D_{kj}(\alpha)=\langle k|\hat{D}(\alpha)|j\rangle$ are then obtained by expanding the exponentials and using $\aa|n\rangle=\sqrt{n}\,|n-1\rangle$ and $\aad|n\rangle=\sqrt{n+1}\,|n+1\rangle$. One finds the standard result
\begin{equation}
D_{kj}(\alpha)
=
e^{-|\alpha|^2/2}
\sqrt{\frac{\min(k,j)!}{\max(k,j)!}}\,
\begin{cases}
\alpha^{\,k-j}\,
L^{(k-j)}_{j}(|\alpha|^2), & k\ge j,\\[2mm]
(-\alpha^*)^{\,j-k}\,
L^{(j-k)}_{k}(|\alpha|^2), & j\ge k,
\end{cases}
\label{eq:Dkj}
\end{equation}
where $L^{(q)}_{n}(x)$ are the generalized Laguerre polynomials. The corresponding transition probabilities between Fock states depend only on $|\alpha|=d$ and can be written in the symmetric form
\begin{equation}
\bigl|D_{kj}(\alpha)\bigr|^2
=
e^{-|\alpha|^2}\,
\frac{\min(k,j)!}{\max(k,j)!}\,
|\alpha|^{2|k-j|}\,
\left[
L^{(|k-j|)}_{\min(k,j)}(|\alpha|^2)
\right]^2.
\label{eq:disp_prob}
\end{equation}

For a general input state $\rho = \sum_{j,m=0}^\infty \rho_{jm}\,|j\rangle\langle m|$, the output populations after the displacement channel are
\begin{equation}
\rho^{(\mathrm{disp})}_{kk}
=
\sum_{j,m=0}^\infty
D_{kj}(\alpha)\,\rho_{jm}\,D_{km}(\alpha)^*,
\end{equation}
and, in particular, for a diagonal input state $\rho=\sum_{j} p_j |j\rangle\langle j|$ one obtains
\begin{equation}
\rho^{(\mathrm{disp})}_{kk}
=
\sum_{j=0}^\infty
p_j\,\bigl|D_{kj}(\alpha)\bigr|^2.
\label{eq:disp_diag_general}
\end{equation}

\subsection{Combination of loss and displacement for initial diagonal states}
\label{sec:loss_disp_combination}

We now combine this with the pure-loss evolution discussed above. After a loss channel of efficiency $\eta(t)=e^{-\Gamma t}$, an initial Fock-diagonal state
$\rho(0)=\sum_{n=0}^\infty p_n |n\rangle\langle n|$ acquires the populations
\begin{equation}
\rho_{jj}(t)
=
\sum_{n=j}^{\infty}
p_n
\binom{n}{j}
\eta(t)^{j}\,[1-\eta(t)]^{\,n-j},
\label{eq:loss_diag_again}
\end{equation}
while all off-diagonal elements vanish. Applying subsequently the displacement channel $\mathcal D_\alpha$, the final populations in Fock state $|k\rangle$ are obtained by inserting
Eq.~\eqref{eq:loss_diag_again} into Eq.~\eqref{eq:disp_diag_general} and using
Eq.~\eqref{eq:disp_prob}, which yields
\begin{equation}
\rho_{kk}^{(\mathrm{loss+disp})}(t)
=
e^{-|\alpha|^2}
\sum_{j=0}^{\infty}
\left(
\sum_{n=j}^{\infty}
p_n
\binom{n}{j}
\eta(t)^{j}\,[1-\eta(t)]^{\,n-j}
\right)
\frac{\min(j,k)!}{\max(j,k)!}\,
|\alpha|^{2|k-j|}\,
\left[
L^{(|k-j|)}_{\min(j,k)}(|\alpha|^2)
\right]^2.
\label{eq:loss_disp_pop}
\end{equation}
This expression gives the probability to measure $|k\rangle$ at time $t$ for an arbitrary Fock-diagonal input state.

\subsection{QFI of Fock-diagonal states in the presence of losses}

We now use the channel decomposition derived above to evaluate the QFI for displacement-amplitude estimation in the presence of phonon loss. For an initial Fock state $\ket{k}$, the combined action of drive and loss leads to a QFI given by a finite, closed-form sum. The same approach extends to arbitrary Fock-diagonal input states, for which the corresponding QFI is generally expressed as an infinite sum.

Consider an initial Fock state $\ket{k}$. At fixed evolution time $t$, the full channel can be written as $\mathcal E_t=\mathcal D_{\alpha(t)}\circ\mathcal L_{\eta(t)}$, with $\eta(t)$ and $\alpha(t)=d(t)e^{i\phi}$ given in Eq.~\eqref{supp:etabeta}. Since $\eta(t)$ is independent of the displacement-drive amplitude, the state after the loss channel, $\rho_{\mathrm{loss}}\equiv\mathcal L_{\eta(t)}[\ket{k}\bra{k}]$, is independent of $d$. The subsequent displacement therefore constitutes the unitary encoding $\rho_d=e^{-id\hat G(\phi)}\rho_{\mathrm{loss}}e^{id\hat G(\phi)}$, where $\phi$ is the fixed displacement direction and $\hat G(\phi)$ is given in Eq.~\eqref{supp:generatorG}.
We may therefore evaluate the QFI of $\rho_\text{loss}$ using the standard machinery for unitarily encoded mixed states.

Since $\ket{k}$ is trivially Fock-diagonal, so is $\rho_\text{loss}$, with populations given by binomial thinning [Eq.~\eqref{eq:loss_diag_again}, specialized to $p_n=\delta_{n,k}$]:
\begin{equation}\label{eq:lambda_binomial}
    \lambda_n \equiv \langle n|\rho_{\mathrm{loss}}|n\rangle = \binom{k}{n}\eta^n(1-\eta)^{k-n}, \qquad n\in\{0,\dots,k\} \fcomma
\end{equation}
and $\lambda_n=0$ for $n>k$. Using the QFI for a unitarily displaced Fock-diagonal state derived in Eq.~\eqref{eq:FQdiag} together with Eq.~\eqref{eq:lambda_binomial}, we have that the sum truncates at $n=k$ (all terms with $n>k$ have $\lambda_n=\lambda_{n+1}=0$ and are dropped), yielding
\begin{equation}\label{eq:FQ_Fock_loss}
    F_Q^{(k)}[d] = 4\sum_{\substack{n=0\\\lambda_n+\lambda_{n+1}>0}}^{k} (n+1)\, \frac{(\lambda_n-\lambda_{n+1})^2}{\lambda_n+\lambda_{n+1}} \fperiod
\end{equation}
This is a finite sum of $k+1$ elementary terms, hence a closed-form expression for any $k$. As expected for a unitary encoding generated by the $d$-independent operator $\hat G(\phi)$, it does not depend on the working point $d$.

Two limiting cases provide useful consistency checks. In the absence of loss, $\eta\rightarrow1$, the state approaches $\ket{k}$ and Eq.~\eqref{eq:FQ_Fock_loss} gives $F_Q^{(k)}[d]=4(2k+1)$, recovering the ideal Fock-state result of Eq.~\eqref{eq:QFI_fock_main}. In the opposite limit, $\eta\rightarrow0$, every initial Fock state relaxes to the vacuum and $F_Q^{(k)}[d]=4$, independently of $k$.

Another interesting limit is obtained for large $k$, for which the distribution $\lambda_n$ changes slowly in $n$.
We can thus approximate $\lambda_n-\lambda_{n+1}\simeq-\partial_n\lambda_n$, $\lambda_n+\lambda_{n+1}\simeq2\lambda_n$ and $n+1\simeq n$ in Eq.~\eqref{eq:FQ_Fock_loss}, obtaining
\begin{equation}\label{suppeq:FQcontinuum}
    F_Q \simeq 2\int dn \;n\, \lambda_n \left(\partial_n \ln\lambda_n\right)^2
    = 2\avg{n\left(\partial_n\ln\lambda_n\right)^2} \fcomma
\end{equation}
where we used $(\partial_n\lambda_n)^2/\lambda_n=\lambda_n(\partial_n\ln\lambda_n)^2$.
Approximating $\lambda_n$ by a Gaussian of mean $\bar{n}$ and variance $\mathrm{Var}[n]$ gives $\partial_n\ln\lambda_n=-(n-\bar{n})/\mathrm{Var}[n]$, and since $\avg{n(n-\bar{n})^2}\simeq\bar{n}\,\mathrm{Var}[n]$ one obtains the simple result
\begin{equation}\label{suppeq:FQfano}
    F_Q \simeq \frac{2\bar{n}}{\mathrm{Var}[n]} \fcomma
\end{equation}
valid for any smooth Fock-diagonal state whose width obeys $1\ll\sqrt{\mathrm{Var}[n]}\ll\bar{n}$.
In this regime, the metrological resource is therefore the number squeezing of the probe, quantified by the Fano factor $\mathrm{Var}[n]/\bar{n}$, rather than its energy.
Heuristically, a pure Fock state has $\mathrm{Var}[n]=0$ and Eq.~\eqref{suppeq:FQfano} diverges, consistently with the unbounded growth of $4(2k+1)$, while a Poissonian distribution has unit Fano factor and gives $F_Q\rightarrow2$, which is indeed the QFI of a phase-randomized coherent state $\bar\rho_\alpha=\frac{1}{2\pi}\int_0^{2\pi}d\theta\,\ket{\alpha e^{i\theta}}\bra{\alpha e^{i\theta}}=\sum_n e^{-|\alpha|^2} |\alpha|^{2n}\ket{n}\bra{n}/n!$.

For the loss-degraded Fock state, Eq.~\eqref{eq:lambda_binomial} gives $\bar{n}=k\eta$ and $\mathrm{Var}[n]=k\eta(1-\eta)$, so that the Fano factor is $1-\eta$ independently of $k$, and
\begin{equation}\label{suppeq:plateau}
    F_Q \;\xrightarrow[k\rightarrow\infty]{}\; \frac{2}{1-\eta} \;=\; \frac{2}{1-e^{-\Gamma t}} \;\simeq\; \frac{2T_1}{t} \fcomma
\end{equation}
where the last expression holds for $t\ll T_1$.
The QFI therefore no longer grows linearly with $k$, but saturates at a value set by the ratio between the lifetime of the mode and the duration of the sensing sequence, no matter how many phonons are in the mode.
As a nontrivial check, at $\eta=1/2$ Eq.~\eqref{suppeq:plateau} predicts $F_Q=4$, and indeed Eq.~\eqref{eq:FQ_Fock_loss} gives $F_Q^{(k)}=4$ exactly, for every $k$ and not only asymptotically.

The approach to the asymptotic value Eq.~\eqref{suppeq:plateau} is not monotonic.
The Gaussian approximation leading to Eq.~\eqref{suppeq:FQfano} requires the binomial distribution to be smooth on both of its sides, i.e.\ $k\eta\gg1$ and $k(1-\eta)\gg1$.
In the small-loss regime of interest the binding condition is the second one, so that the relevant expansion parameter is the inverse of the mean number of phonons lost during the sequence, $\bar\nu=k(1-\eta)$.
When $\bar\nu\gg1$ the leading correction is
\begin{equation}\label{suppeq:overshoot}
    F_Q^{(k)} \simeq \dfrac{2}{1-\eta}\left[1+\dfrac{1}{4\bar\nu}\right] \fcomma
\end{equation}
so that the plateau is approached from above, as $1/k$.
For $\bar\nu\lesssim1$, on the other hand, the expansion breaks down.
In this regime it is convenient to count instead the number of lost phonons $m=k-n$, which for $\eta\rightarrow1$ follows a Poisson distribution of mean $\bar\nu$.
The ratio between $F_Q^{(k)}$ and the plateau value Eq.~\eqref{suppeq:plateau} is then a function of $\bar\nu$ alone, which we find exceeds unity by at most $\sim 8\%$ and is maximal when $\bar\nu\simeq 1.7$ phonons are lost on average.
The overshoot is therefore modest, and the plateau value remains a good estimate of the sensitivity.
For $\eta=0.9$, for instance, $F_Q$ peaks at $21.5$ for $k=15$ ($\bar\nu=1.5$) and has fallen to $20.4$ at $k=100$ ($\bar\nu=10$), against an asymptotic value of $20$.

Finally, the drive need not be constant in time. For a fixed pulse shape, we write $\Omega(\tau)=\Omega f(\tau)$, where $f(\tau)$ is a known dimensionless envelope and $\Omega$ sets its overall real amplitude. At fixed evolution time $t$, Eq.~\eqref{supp:etabeta} then gives a linear relation between the displacement amplitude and the drive amplitude,
\begin{equation}
d(t)=c_f(t)\,\Omega \;, \qquad c_f(t)\equiv\int_0^t d\tau\,e^{-\Gamma(t-\tau)/2}f(\tau) \fperiod
\end{equation}
The QFI therefore transforms according to the standard reparametrization rule,
\begin{equation}\label{eq:Omega_from_d_QFI}
F_Q^{(k)}[\Omega]=\left(\frac{\partial d}{\partial\Omega}\right)^2F_Q^{(k)}[d]=c_f^2(t)\,F_Q^{(k)}[d] \fperiod
\end{equation}
Since the total evolution time $t$, and hence $\eta(t)$ and $\rho_\text{loss}$, are fixed independently of $\Omega$, Eq.~\eqref{eq:FQ_Fock_loss} gives the QFI of the output state with respect to the displacement amplitude $d$, while Eq.~\eqref{eq:Omega_from_d_QFI} expresses the same information with respect to the experimentally controlled drive amplitude $\Omega$. For a constant drive, $f(\tau)=1$ and $c_f(t)=2\left(1-e^{-\Gamma t/2}\right)/\Gamma$.

\vspace{3mm}

The same derivation extends directly to an arbitrary Fock-diagonal input state $\rho(0)=\sum_n p_n\ket{n}\bra{n}$. Since the loss channel preserves Fock diagonality, the state $\rho_\text{loss}$ is fully characterized by the loss-transformed populations $\lambda_n=\langle n|\rho_\text{loss}|n\rangle$, obtained from the binomial-thinning relation derived above. The QFI then follows directly from Eq.~\eqref{eq:FQdiag}, since its derivation relies only on the Fock-diagonal form of $\rho_\text{loss}$ and the nearest-neighbor coupling generated by $\hat G(\phi)$. 
For an initial Fock state the loss-transformed distribution has finite support and the sum reduces to Eq.~\eqref{eq:FQ_Fock_loss}, whereas for a general Fock-diagonal input it may contain infinitely many nonzero terms.

The reparametrization to the physical drive amplitude, Eq.~\eqref{eq:Omega_from_d_QFI}, is independent of the input state and therefore applies unchanged. For a fixed pulse shape and evolution time, it depends only on the linear relation $d(t)=c_f(t)\Omega$. The only distinction from the initial Fock-state case is the support of the loss-transformed distribution: if the input populations $p_n$ have finite support, the QFI reduces to a finite sum as in Eq.~\eqref{eq:FQ_Fock_loss}, whereas for distributions with infinite support, such as thermal or Poissonian states, the expression remains exact but generally involves an infinite series. In the no-loss limit $\eta\rightarrow1$, the loss-transformed populations approach the input populations and the QFI reduces to that of the corresponding Fock-diagonal input state. In the opposite limit $\eta\rightarrow0$, the loss channel maps any normalized input state to the vacuum, yielding $F_Q[d]\rightarrow4$, independently of the initial phonon-number distribution.

\subsection{Measurement optimality in the presence of losses}
\label{supp:nonoptRPNLoss}
The QFI in Eq.~\eqref{eq:FQ_Fock_loss} is locally attained by a projective measurement in the eigenbasis of the symmetric logarithmic derivative (SLD) $\hat L_d$, defined by
\begin{equation}
\partial_d\rho_d=\frac{1}{2}\left(\hat L_d\rho_d+\rho_d\hat L_d\right) \fperiod
\end{equation}
At the reference point $d=0$, the state is $\rho_\text{loss}=\sum_n\lambda_n\ket{n}\bra{n}$ and $\partial_d\rho_d|_{d=0}=-i[\hat G(\phi),\rho_\text{loss}]$. The SLD matrix elements are therefore
\begin{equation}
\langle n|\hat L_0|m\rangle=2i\,\frac{\lambda_n-\lambda_m}{\lambda_n+\lambda_m}\langle n|\hat G(\phi)|m\rangle \fcomma
\end{equation}
for $\lambda_n+\lambda_m>0$. Since $\hat G(\phi)$ couples only neighboring Fock states, this reduces for the loss-transformed Fock state of Eq.~\eqref{eq:lambda_binomial} to
\begin{equation}\label{eq:SLD_general_loss}
\hat L_0=2\sum_{n=0}^{k}\frac{(n+1)(1-\eta)-(k-n)\eta}{(n+1)(1-\eta)+(k-n)\eta}\sqrt{n+1}\left(e^{-i\phi}\ket{n}\bra{n+1}+e^{i\phi}\ket{n+1}\bra{n}\right) \fcomma
\end{equation}
where $\lambda_{k+1}\equiv0$. For $0<\eta<1$, all terms in Eq.~\eqref{eq:SLD_general_loss} are well defined. At the limiting values $\eta=0,1$, matrix elements acting entirely within the kernel of the state are arbitrary and may be set to zero. Direct evaluation gives
\begin{equation}
\Tr[\rho_\text{loss}\hat L_0^2]=4\sum_{\substack{n=0\\\lambda_n+\lambda_{n+1}>0}}^{k}(n+1)\frac{(\lambda_{n+1}-\lambda_n)^2}{\lambda_{n+1}+\lambda_n}=F_Q^{(k)}[d] \fcomma
\end{equation}
consistent with Eq.~\eqref{eq:FQ_Fock_loss}.

For finite $d$, the SLD follows from the same unitary covariance as the encoded state,
\begin{equation}
\hat L_d=e^{-id\hat G(\phi)}\hat L_0e^{id\hat G(\phi)}
=\hat D(de^{i\phi})\hat L_0\hat D^\dagger(de^{i\phi}) \fperiod
\end{equation}

In the lossless limit $\eta\rightarrow1$, $\rho_\text{loss}\rightarrow\ket{k}\bra{k}$. A choice of SLD that vanishes within the kernel of the pure state is
\begin{equation}
\hat L_0=-2i[\hat G(\phi),\ket{k}\bra{k}] \fcomma
\end{equation}
or explicitly
\begin{equation}\label{eq:SLD_pure_limit}
\hat L_0=2\sqrt{k+1}\left(e^{-i\phi}\ket{k}\bra{k+1}+e^{i\phi}\ket{k+1}\bra{k}\right)-2\sqrt{k}\left(e^{-i\phi}\ket{k-1}\bra{k}+e^{i\phi}\ket{k}\bra{k-1}\right) \fcomma
\end{equation}
with the terms containing $\ket{k-1}$ absent for $k=0$. This gives $\Tr[\ket{k}\bra{k}\hat L_0^2]=4(2k+1)$, as required. Equation~\eqref{eq:SLD_general_loss} approaches the same support-relevant SLD in the limit $\eta\rightarrow1$; additional matrix elements connecting states entirely outside the support of $\ket{k}$ are immaterial and reflect the non-uniqueness of the SLD on the kernel of a rank-deficient state.

In contrast to the pure-Fock-state case, number-resolved detection does not generally saturate the QFI once phonon loss renders the probe mixed. This does not follow merely from the form of the SLD, since its eigenbasis need not be the unique optimal measurement. Rather, it follows from direct comparison between the classical FI for Fock-basis detection in Eq.~\eqref{eq:FCdiag} and the mixed-state QFI in Eq.~\eqref{eq:FQ_Fock_loss}, which are not equal in general for $0<\eta<1$, although equality can occur at particular parameter values.

\clearpage
\newpage

\subsection{Saturation of the maximum FI}

In this section we explain the saturation of the measured Fisher information observed in Fig.~\ref{fig:3} by providing a minimal model which takes into account only the measured Fock-state preparation fidelities combined with phonon loss during the sensing sequence.

\begin{figure}
    \centering
    \includegraphics[width=\linewidth]{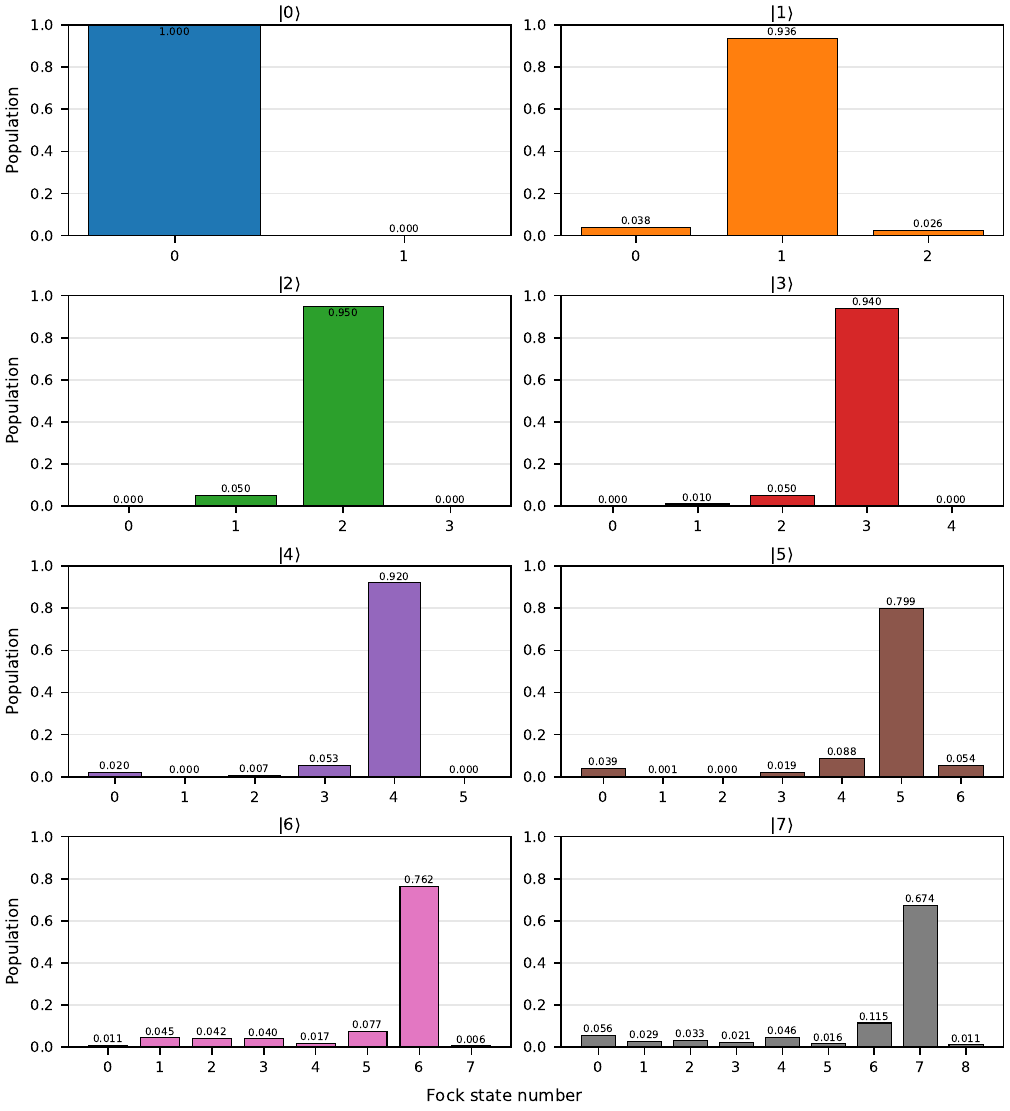}
    \caption{\textbf{Measured Fock state distributions for the initial states.} For each of the experimentally prepared states, we show the Fock state occupations as obtained from RPN measurement. The tallest bar of each histogram corresponds to the fidelities $\mathcal{F}_k$.}
    \label{suppfig:measFockDist}
\end{figure}

Let us consider a probe state that is Fock-diagonal, $\rho= \sum_n \lambda_n \ket{n}\bra{n}$, as can result from imperfections in the optimal-control pulse and from phonon decay during preparation.
For such a state, the QFI for displacement sensing is given by Eq.~\eqref{eq:FQdiag}.
In order to construct an ansatz for $\rho$, let $\mathcal{F}_k=\langle k|\rho|k\rangle$ denote the measured population of the target state $\ket{k}$ at zero displacement, i.e. its fidelity, see Fig.~\ref{suppfig:measFockDist}.
Since the dominant preparation imperfection is energy relaxation, we choose to assign all the missing population to the state with one phonon less,
\begin{equation}\label{suppeq:prepmodel}
    \rho = \mathcal{F}_k\ket{k}\bra{k} + (1-\mathcal{F}_k)\ket{k-1}\bra{k-1} \fperiod
\end{equation}

The displacement pulse then acts through the channel $\mathcal E_{t}=\mathcal D_{\alpha}\circ\mathcal L_{\eta}$ of Eq.~\eqref{supp:chDec}.
Only the loss part modifies the populations according to the binomial thinning Eq.~\eqref{eq:loss_diag_again}, so that the state before the unitary displacement is Fock-diagonal with
\begin{equation}\label{suppeq:lossypops}
    \lambda_n^{(k)} = \mathcal{F}_k\,B_{n|k}(\eta) + (1-\mathcal{F}_k)\,B_{n|k-1}(\eta) \fcomma
    \qquad
    B_{n|m}(\eta)=\binom{m}{n}\eta^n(1-\eta)^{m-n} \fcomma
\end{equation}
and $\eta=e^{-\Gamma t}$.
Inserting Eq.~\eqref{suppeq:lossypops} into Eq.~\eqref{eq:FQdiag} gives the QFI of the probe actually used in the experiment.

We can now compare different scenarios, which are summarized in Tab.~\ref{supptab:plateau} and in Fig.~\ref{fig:3} of the main text.
First, we recall that Fock state $\ket{k}$ has QFI $4(2k+1)$ for displacement sensing (column ``Ideal'' of Tab.~\ref{supptab:plateau} and orange line in Fig.~\ref{fig:3}). 
If losses occur during the displacement we can follow the argument presented around Eq.~\eqref{suppeq:overshoot} to get an idea of the sensitivity as a function of $k$. Taking $\eta=0.97$ ($\Gamma=1/\unit{196}{\micro\second}$, $t=\unit{6}{\micro\second}$) we expect a sensitivity plateau with $F_Q\simeq 2/(1-\eta)\simeq 67$ approached from above after a maximum $F_Q\simeq 72$ at $k^\ast\simeq 54$. 
Both lie far above the range explored here, $k\leq7$, which is why the corresponding curve in Fig.~\ref{fig:3} is still rising over the whole measured interval (column ``Ideal prep, lossy disp'' of Tab.~\ref{supptab:plateau} and green curve in Fig.~\ref{fig:3}).
Phonon loss during the sequence is therefore not sufficient to explain the saturation observed experimentally at $\bar{n}\gtrsim 3$.
However, taking into account imperfect initial state preparation with the model Eq.~\eqref{suppeq:prepmodel} shows a saturation already for $\bar{n}\gtrsim 4$, in better agreement with the one observed in the experiment (column ``Prep error and lossy disp'' of Tab.~\ref{supptab:plateau} and red curve in Fig.~\ref{fig:3}).
Still, the plateau value $F_Q\simeq25$ lies above the experimentally observed $F_\text{max}\simeq20$.
This gap can be attributed to calibration imperfections, to the finite resolution of the RPN measurement, and to the fact that, as discussed in Sec.~\ref{supp:nonoptRPNLoss}, number-counting measurements are not necessarily optimal for mixed probe states, so that the classical FI need not saturate the QFI.

\begingroup
\setlength{\tabcolsep}{8pt}
\renewcommand{\arraystretch}{1.2}
\begin{table}[t]
\caption{\justifying Displacement sensitivity of the probe states, evaluated from the minimal model of Eq.~\eqref{suppeq:prepmodel} with the measured preparation fidelities $\mathcal{F}_k$ and a loss channel of transmissivity $\eta=e^{-\Gamma t}=0.97$.}
\label{supptab:plateau}
\begin{tabular}{c|ccc|cc|cc|cc}
 & \multicolumn{3}{|c|}{Measured} & \multicolumn{2}{|c|}{Ideal} & \multicolumn{2}{|c|}{Ideal prep, lossy disp} & \multicolumn{2}{|c}{Prep error and lossy disp} \\
\hline
\makecell{Target \\Fock $k$} & Fidelity $\mathcal{F}_k$ & $\bar{n}$ & $F_{\max}$ & $\bar{n}$ & \makecell{QFI $4(2k+1)$} & $\bar{n}$ & $F_{\max}$ & $\bar{n}$ & $F_{\max}$ \\
\hline
0 & 1.000 & 0.00 &  4.4(3) & 0 &  4 & 0.00 &  4.0 & 0.00 &  4.0 \\
1 & 0.936 & 0.99 & 11.4(5) & 1 & 12 & 0.97 & 11.3 & 0.91 &  9.9 \\
2 & 0.950 & 1.90 & 15.0(6) & 2 & 20 & 1.94 & 17.8 & 1.89 & 16.1 \\
3 & 0.940 & 2.82 & 19.5(8) & 3 & 28 & 2.91 & 23.5 & 2.85 & 21.0 \\
4 & 0.920 & 3.61 & 20.4(8) & 4 & 36 & 3.88 & 28.6 & 3.80 & 24.7 \\
5 & 0.799 & 4.49 & 20.1(8) & 5 & 44 & 4.85 & 33.1 & 4.66 & 23.5 \\
6 & 0.762 & 5.04 & 19.4(8) & 6 & 52 & 5.82 & 37.1 & 5.59 & 25.2 \\
7 & 0.674 & 5.73 & 19.4(8) & 7 & 60 & 6.79 & 40.7 & 6.47 & 24.9 \\
\end{tabular}
\end{table}
\endgroup

\clearpage
\newpage

\clearpage
\newpage

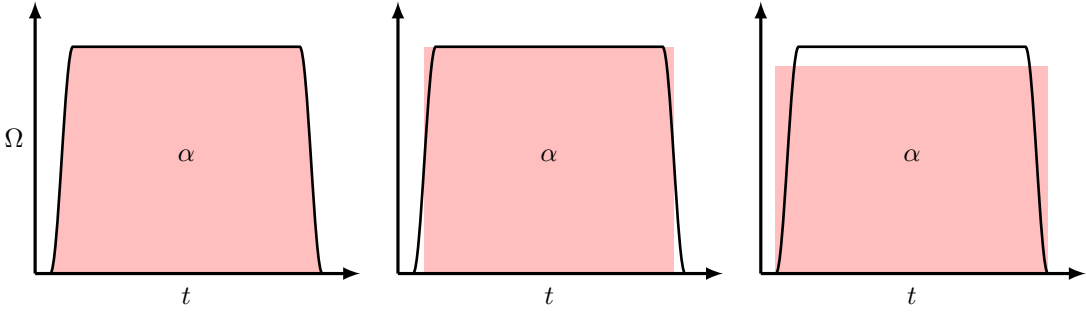
\begin{figure}[t]
    \centering
    \input{effective_drive_amplitude.tikz}
    \caption{
        Comparison of the displacement relation to the drive strength.
        The red area depicts how $d(t)=\int_{0}^{t}d\tau\,\Omega(\tau)$ is obtained, and shows the effective signal that created this displacement.
        The area stays the same between all panels.
        \textbf{a)} true, time-dependent drive of Eq.~\eqref{supp:pulseshape}.
        \textbf{b)} effective duration $t_{d,\text{eff}}$ at fixed drive strength $\Omega$ [Eq.~\eqref{supp:effintegral}].
        \textbf{c)} effective drive strength $\Omega_{\text{eff}}$ at fixed total duration $t_d$ [Eq.~\eqref{supp:effstrength}].
        }
    \label{suppfig:effectiveDrive}
\end{figure}

\section{Displacement drive calibration}

To calibrate the strength of the displacement we apply in the experiment, we consider the following procedure.
We start with the oscillator in the ground state $\ket{0}$ and apply the resonant displacement drive for a total duration $t_d=\unit{6}{\micro\second}$, identical to the duration of the drive whose amplitude we want to sense. 
The drive consists of a constant-amplitude flat-top segment of duration $t_c=\unit{5}{\micro\second}$ and two ramps of duration $t_r=\unit{0.5}{\micro\second}$ each, giving a total duration $t_d=t_c+2t_r$; see Fig.~\ref{suppfig:effectiveDrive}a.
The ramps are modeled as half-cosine functions of amplitude $\Omega$, such that
\begin{equation}\label{supp:pulseshape}
    \Omega(t) =   \begin{cases}
    \dfrac{\Omega}{2}\left[1 - \cos(\pi t/t_r)\right], & 0 \leq t \leq t_r \\[1mm]
    \Omega, & t_r \leq t \leq t_r+t_c \\[1mm]
    \dfrac{\Omega}{2}\left[1 + \cos\big(\pi (t-t_r-t_c)/t_r\big)\right], & t_r+t_c \leq t \leq t_d
  \end{cases} \;.
\end{equation}
According to the Hamiltonian Eq.~\eqref{eq:dispH}, this time-dependent envelope results, in the absence of losses, in a displacement $\alpha(t)=d(t)e^{i\phi}$, with $d(t)=\int_{0}^{t}d\tau\,\Omega(\tau)$ and $\Omega(\tau)$ the actual time-dependent drive strength given by Eq.~\eqref{supp:pulseshape}.
Using $\int_{0}^1 dx\, [1 \pm \cos(\pi x)]/2=1/2$, each ramp segment contributes exactly half of a flat segment of duration $t_r$, so that
\begin{equation}\label{supp:effintegral}
    \int_{0}^{t_d} d\tau\,\Omega(\tau) = \Omega\,(t_c + t_r) \equiv \Omega\, t_{d,\text{eff}} \;,
\end{equation}
defining the effective pulse duration $t_{d,\text{eff}}\equiv (t_c+t_r)$, i.e., the flat-top duration extended by a single ramp width to account for the two half-weighted edges, see Fig.~\ref{suppfig:effectiveDrive}b. In the absence of losses one thus has
\begin{equation}\label{supp:OmegaFromBeta}
    d(t_d) = \Omega \, t_{d,\text{eff}} \;,
\end{equation}
which can be inverted to convert a measured $d$ into the corresponding drive amplitude $\Omega$.
Being this relation linear, error propagation via the Fisher-information chain rule gives the sensitivity for $\Omega$ in terms of that for $d$,
\begin{equation}\label{supp:FisherOmegaBeta}
    F_{Q/C}[\Omega] = \left( \dfrac{\partial d}{\partial \Omega}\right)^{\!2} F_{Q/C}[d] =  (t_c + t_r)^2\, F_{Q/C}[d] \;.
\end{equation}

Equation~\eqref{supp:effintegral} admits an equivalent interpretation, see Fig.~\ref{suppfig:effectiveDrive}c. Instead of defining an effective duration for fixed amplitude $\Omega$, one can keep the pulse duration fixed to $t_d$ and rescale its amplitude by defining the effective drive strength $\Omega_{\text{eff}} \equiv \Omega\, t_{d,\text{eff}}/t_d$ such that
\begin{equation}\label{supp:effstrength}
    d(t_d) = \Omega \, t_{d,\text{eff}} = \Omega_{\text{eff}} \, t_d \;.
\end{equation}

The above discussion considers an ideal, unitary displacement.
With phonon losses present, the dynamics of the system follows the master equation~\eqref{eq:master}, and the resulting state can be computed using the channel decomposition Eq.~\eqref{supp:chDec}, $\mathcal{E}_t = \mathcal D_{\alpha(t)} \circ \mathcal L_{\eta(t)}$, with $\eta(t)$ and $\alpha(t)$ given in Eq.~\eqref{supp:etabeta}. Starting from vacuum, this results in the coherent state $\ket{\alpha(t)}$, with displacement amplitude $d(t)$. Although Eq.~\eqref{supp:etabeta} can be evaluated analytically for our pulse shape, a simpler approximate result is obtained by replacing the smooth-edged pulse by a constant drive of strength $\Omega_\text{eff}$ applied over the total elapsed time $t_d$. This gives Eq.~\eqref{eq:dToOmega} in the main text, namely
\begin{equation}\label{supp:betafull}
d(t_d) =  \frac{2\Omega_\text{eff}}{\Gamma} \left( 1 - e^{-\Gamma t_d /2} \right) = \frac{2\Omega}{\Gamma} \dfrac{t_{d,\text{eff}}}{t_d} \left(1 - e^{-\Gamma t_d /2} \right) \fperiod
\end{equation}
Replacing the actual smooth-edged pulse by this effective constant-amplitude pulse agrees exactly in the lossless limit [Eq.~\eqref{supp:effstrength}] and remains an excellent approximation when the ramp duration is short compared with both the total pulse duration and the damping timescale, as is the case here ($t_r/t_d\approx8\%$ and $\Gamma t_r\ll1$).

\begin{figure}
    \centering
    \includegraphics[width=0.4\linewidth]{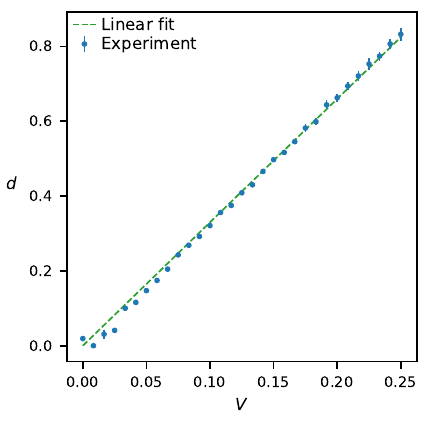}
    \caption{\textbf{Calibration of the drive strength.} Phase-space displacement amplitude $d=|\alpha|$ extracted from Poissonian fits to the displaced-vacuum phonon-number distributions as a function of the displacement-drive setting. Points show the experimental values with error bars, and the dashed line is a linear fit used to convert the experimental control amplitude to the calibrated phase-space displacement.}
    \label{suppfig:driveCalib}
\end{figure}

The coherent state $\ket{\alpha(t_d)}$ obtained by displacing the vacuum has a Poissonian phonon-number distribution,
\begin{equation}
P(n|d)=|\langle n|\alpha(t_d)\rangle|^2=e^{-d^2}\frac{d^{2n}}{n!} \fcomma
\end{equation}
with mean occupation $\bar n=d^2=|\alpha(t_d)|^2$. Experimentally, for each microwave-generator voltage $V_i$, we apply the displacement pulse, reconstruct the phonon-number distribution, and fit it with a Poisson distribution to extract the mean occupation $\bar n_i$. A small drive-independent excess population is present even at zero applied drive and is accounted for phenomenologically by an offset $\bar n_0$, such that $\bar n_{\rm fit}(V)=d^2(V)+\bar n_0$. The displacement amplitude generated by the drive is therefore $d(V)=\sqrt{\bar n_{\rm fit}(V)-\bar n_0}$.
Since the drive amplitude is proportional to the microwave-generator voltage, we fit the extracted displacement amplitudes $d_i$ linearly as a function of $V_i$, as shown in Fig.~\ref{suppfig:driveCalib}. Using Eq.~\eqref{supp:betafull}, the corresponding drive amplitude $\Omega(V)$ is
\begin{equation}\label{supp:OmegaCalibration}
\Omega(V)=\dfrac{\Gamma t_d}{2t_{d,\mathrm{eff}}\left(1-e^{-\Gamma t_d/2}\right)}\,d(V)
=\dfrac{\Gamma t_d}{2t_{d,\mathrm{eff}}\left(1-e^{-\Gamma t_d/2}\right)}
\sqrt{\bar n_{\rm fit}(V)-\bar n_0} \fperiod
\end{equation}
This provides the calibration between the experimental control voltage $V$, the displacement amplitude $d$, and the corresponding drive amplitude $\Omega$ used in the main text.

\clearpage
\newpage

\section{Sampling Fock distributions}

Consider a bosonic mode that is prepared and measured $M$ times, and let $p_n$ denote the (unknown) true population of Fock state $\ket{n}$. We compare two strategies for extracting $\{p_n\}$ from the data:
\begin{enumerate}
    \item \emph{Direct} measurement of the Fock distribution via the POVM $\{ \ket{n}\bra{n} \}_n$.
    \item \emph{Indirect} measurement via a Jaynes-Cummings (JC) interaction of variable duration $t$ between the mode and an initially excited qubit, followed by qubit readout in the $\sigma_z$ basis. This corresponds to the POVM $\{\hat E_e(t),\hat E_g(t)\}$ with
    \begin{equation}
        \hat E_e(t)=\sum_n \cos^2 \big(gt\sqrt{n+1}\big)\, \ket{n}\bra{n}, \qquad \hat E_g(t)=\mathds{1}-\hat E_e(t) \;,
    \end{equation}
    where $g$ is the qubit-phonon coupling strength. This is the resonant phonon-number (RPN) measurement used throughout this work, in which $t$ is swept over a discrete set of interaction times.
\end{enumerate}
The form of $\hat E_e(t)$ follows from the JC dynamics: if the qubit starts in $\ket{e}$ and the mode in $\ket{n}$, coherent Rabi oscillation at rate $2g\sqrt{n+1}$ within the two-dimensional JC manifold spanned by
$\{\ket{e,n},\ket{g,n+1}\}$ gives an excited-state return probability $\cos^2(gt\sqrt{n+1})$; for a general mixed mode state $\rho=\sum_n p_n\ket{n}\bra{n}$, the qubit excited-state probability at time $t$ is therefore $P_e(t)=\Tr[\hat E_e(t)\rho]=\sum_n p_n \cos^2(gt\sqrt{n+1})$.

\nosubsubsection{Method 1: direct sampling}
The $M$ measurements directly sample the Fock probability distribution. We define the empirical frequency $f_n=\#_n/M$, with $\#_n$ the number of outcomes registering Fock state $n$. Since $\#_n$ follows a Binomial distribution with success probability $p_n$ over $M$ trials, the sampling variance of $f_n$ is
\begin{equation}
    \var{f_n} = \dfrac{p_n(1-p_n)}{M} \leq \dfrac{1}{4M} \;.
    \label{eq:varf_direct}
\end{equation}

\nosubsubsection{Method 2: RPN measurement and linear inversion}
Method 2 instead measures linear combinations of the Fock probabilities $p_n$, so that their extraction requires solving a linear inverse problem, which we do by least-squares fitting.

In an RPN measurement, $m=M/n_T$ qubit measurements are taken at each of $n_T$ interaction times $\{t_1,\dots,t_{n_T}\}$, for a total of $M$ measurements. At each $t_i$, the qubit outcome follows a Bernoulli distribution with mean $P_e(t_i)$; the corresponding empirical frequency $F_e(t_i)=\#_e(t_i)/m$ has variance
\begin{equation}
    \var{F_e(t_i)} = \dfrac{P_e(t_i)[1-P_e(t_i)]}{m} \leq \dfrac{1}{4m} \;.
    \label{eq:varFe}
\end{equation}
Approximating $P_e(t)\approx F_e(t)$, we solve the linear system
\begin{equation}
    P_e(t) = \sum_{n=0}^{n_\text{max}} p_n\, K_n(t) \;, \qquad K_n(t)=\cos^2 \big(gt\sqrt{n+1}\big) \;,
    \label{eq:linsystem}
\end{equation}
truncated at a cutoff $n_\text{max}$; $K_n(t)$ may alternatively be obtained numerically from an RPN simulation including decoherence.

In matrix form, the ideal relation between the true probabilities is $\vec P = A\vec p$, with $\vec P=(P_e(t_1),\dots,P_e(t_{n_T}))^{\top}$, $\vec p=(p_0,\dots,p_{n_\text{max}})^{\top}$, and $A_{tn}=K_n(t)$. The measured frequencies obey $\vec F = A\vec p + \vec\epsilon$, where $\vec\epsilon$ has zero mean and uncorrelated entries with
$\var{\epsilon_i}=P_e(t_i)[1-P_e(t_i)]/m\leq 1/(4m) = n_T/4M$ [Eq.~\eqref{eq:varFe}].

Ignoring for the moment the physical constraints $p_n\geq 0$ and $\sum_n p_n=1$ (which can be imposed \emph{a posteriori} via constrained least-squares, e.g.\ using slack variables), the ordinary least-squares estimator is
\begin{equation}
    \vec f_{LS} = (A^\top A)^{-1} A^\top \vec F \;,
    \label{eq:LSestimator}
\end{equation}
valid provided $A$ has full column rank, which requires $n_T\geq n_\text{max}+1$ and a choice of interaction times for which the columns of $A$ are linearly independent.
Substituting $\vec F = A\vec p + \vec\epsilon$ gives $\vec f_{LS} = \vec p + (A^\top A)^{-1}A^\top \vec\epsilon$, and since $\Sigma\equiv\cov{\vec\epsilon}{\vec\epsilon}
=\operatorname{diag}\left[
\frac{P_e(t_i)[1-P_e(t_i)]}{m}
\right]
\preceq \mathds{1}/4m$,
\begin{equation}
    \cov{\vec f_{LS}}{\vec f_{LS}} = (A^\top A)^{-1}A^\top\Sigma A(A^\top A)^{-1}
    \preceq
    \dfrac{n_T}{4M}(A^\top A)^{-1}.
    \label{eq:covLS}
\end{equation}
Writing $B\equiv (A^\top A)^{-1}A^\top$, so that $f_n = \sum_t B_{nt} F_e(t)$, and assuming independent measurements at different times, Eq.~\eqref{eq:covLS} implies
\begin{equation}
    \var{f_n} = \sum_i B_{ni}^2\,\var{F_e(t_i)} \leq \dfrac{n_T}{4M} \big[(A^\top A)^{-1}\big]_{nn} \;,
    \label{eq:varf_indirect}
\end{equation}
using $\sum_t B_{nt}^2 = (BB^\top)_{nn}=[(A^\top A)^{-1}]_{nn}$. Since the diagonal entries of $(A^\top A)^{-1}$ are bounded by the extremal eigenvalues of $A^\top A$, 
\begin{equation}
    \var{f_n} \leq \dfrac{n_T}{4M\,\lambda_{\min}(A^\top A)} \fperiod
    \label{eq:varf_bounds}
\end{equation}

\vspace{2mm}
\textbf{RPN frequency-resolution consideration.} Since the basis functions $K_n(t)=\cos^2(gt\sqrt{n+1})=\big[1+\cos(2gt\sqrt{n+1})\big]/2$ oscillate at frequency $\omega_n = 2g\sqrt{n+1}$, resolving neighboring Fock states requires resolving the frequency spacing
\begin{equation}
    \Delta\omega_n = \omega_{n+1}-\omega_n = 2g\big(\sqrt{n+2}-\sqrt{n+1}\big) \approx \dfrac{g}{\sqrt{n+1}} \;,
\end{equation}
which shrinks with increasing $n$. 
A simple Fourier-resolution estimate for resolving up to $n_\text{max}$ is obtained from requiring to resolve the smallest spacing $\Delta\omega_{n_\text{max}}$, i.e.\ a total interrogation window
\begin{equation}
    T_\text{max} \gtrsim \dfrac{2\pi}{\Delta\omega_{n_\text{max}}} \approx \dfrac{2\pi\sqrt{n_\text{max}+1}}{g} \;.
\end{equation}
Sampling this window with a fixed time step $\Delta t$ requires $n_T = T_\text{max}/\Delta t$ points, so that, inverting,
\begin{equation}
    n_T \approx \dfrac{2\pi\sqrt{n_\text{max}+1}}{g\,\Delta t}
    \qquad\Longleftrightarrow\qquad
    n_\text{max} \approx \left(\dfrac{g\,T_\text{max}}{2\pi}\right)^{2} - 1 \;.
    \label{eq:nmax_scaling}
\end{equation}

\nosubsubsection{Effective sample number}
We can now directly compare the statistical efficiency of the two methods. Method 1 gives $\var{f_n}\leq 1/4M$ [Eq.~\eqref{eq:varf_direct}], while Method 2 gives $\var{f_n}\leq (n_T/4M)/\lambda_{\min}(A^\top A)$ [Eq.~\eqref{eq:varf_bounds}]. Equating the two right-hand sides, 
a conservative worst-case effective sample number for an RPN measurement using $M$ total measurements is obtained by matching these variance upper bounds, 
\begin{equation}
    M^\ast = M\, \dfrac{\lambda_{\min}(A^\top A)}{n_T} \;.
    \label{eq:effShotsM}
\end{equation}
As a representative example, for $n_T=500$ equally-spaced measurement times and a truncation $n_\text{max}=10$, we find $\lambda_{\min}(A^\top A)/n_T \approx 0.088$, i.e.\ $M^\ast \approx 0.09\,M$: in this case, the worst-case variance bound indicates that RPN requires roughly $10$ times more measurements than direct Fock-state sampling to reach the same $\var{f_n}$. This factor quantifies the overhead intrinsic to reconstructing the phonon-number distribution indirectly through the JC-mediated qubit readout, and is used in the main text to relate the number of RPN repetitions to an equivalent direct-sampling statistics.

\end{document}

%% file: effective_drive_amplitude.tikz
\begin{tikzpicture}[
    x=1cm, y=1cm,
    >={latex},
    line/.style={line width=1.0pt},
    aline/.style={line width=1.2pt},
    nlabel/.style={font=\normalsize},
]

\useasboundingbox (0,0) rectangle (14.5,4.2);

\def\yStart{0.5}      
\def\yAmplitude{3.0}  
\def\yEnd{\yAmplitude*1.2}
\def\ylabel{-0.3} 

\def\xA{0.5}  
\def\xB{5.3}  
\def\xC{10.1}  

\def\xStart{0.2}  
\def\xEdge{0.3}  
\def\xPulse{3.0}  
\def\xEnd{0.5}    
\def\xlabel{-0.28} 

\pgfmathsetmacro{\xAxisLen}{\xStart+\xEdge+\xPulse+\xEdge+\xEnd}

\fill[red!25, draw=none]
    (\xA+\xStart,\yStart)
    cos (\xA+\xStart+0.5*\xEdge,\yStart+0.5*\yAmplitude)
    sin (\xA+\xStart+\xEdge,\yStart+\yAmplitude)
    -- (\xA+\xStart+\xEdge+\xPulse,\yStart+\yAmplitude)
    cos (\xA+\xStart+\xEdge+\xPulse+0.5*\xEdge,\yStart+0.5*\yAmplitude)
    sin (\xA+\xStart+\xEdge+\xPulse+\xEdge,\yStart) -- cycle;

\draw[aline,->] (\xA,\yStart) -- (\xA+\xAxisLen,\yStart);
\draw[aline,->] (\xA,\yStart) -- (\xA,\yStart+\yEnd);

\draw[line]
    (\xA+\xStart,\yStart)
    cos (\xA+\xStart+0.5*\xEdge,\yStart+0.5*\yAmplitude)
    sin (\xA+\xStart+\xEdge,\yStart+\yAmplitude)
    -- (\xA+\xStart+\xEdge+\xPulse,\yStart+\yAmplitude)
    cos (\xA+\xStart+\xEdge+\xPulse+0.5*\xEdge,\yStart+0.5*\yAmplitude)
    sin (\xA+\xStart+\xEdge+\xPulse+\xEdge,\yStart);

\node[nlabel] at (\xA+\xlabel,\yStart+0.5*\yEnd) {$\Om$};
\node[nlabel] at (\xA+\xStart+\xEdge+0.5*\xPulse,\yStart+0.52*\yAmplitude) {$\displacement$};
\node[nlabel] at (\xA+\xStart+\xEdge+0.5*\xPulse,\yStart+\ylabel) {$t$};

\fill[red!25, draw=none]
    (\xB+\xStart+0.5*\xEdge, \yStart) rectangle
    (\xB+\xStart+\xEdge+\xPulse+0.5*\xEdge, \yStart+\yAmplitude);

\draw[aline,->] (\xB,\yStart) -- (\xB+\xAxisLen,\yStart);
\draw[aline,->] (\xB,\yStart) -- (\xB,\yStart+\yEnd);

\draw[line]
    (\xB+\xStart,\yStart)
    cos (\xB+\xStart+0.5*\xEdge,\yStart+0.5*\yAmplitude)
    sin (\xB+\xStart+\xEdge,\yStart+\yAmplitude)
    -- (\xB+\xStart+\xEdge+\xPulse,\yStart+\yAmplitude)
    cos (\xB+\xStart+\xEdge+\xPulse+0.5*\xEdge,\yStart+0.5*\yAmplitude)
    sin (\xB+\xStart+\xEdge+\xPulse+\xEdge,\yStart);

\node[nlabel] at (\xB+\xStart+\xEdge+0.5*\xPulse,\yStart+0.52*\yAmplitude) {$\displacement$};
\node[nlabel] at (\xB+\xStart+\xEdge+0.5*\xPulse,\yStart+\ylabel) {$t$};

\fill[red!25, draw=none]
    (\xC+\xStart, \yStart) rectangle
    (\xC+\xStart+\xEdge+\xPulse+\xEdge, {\yStart+\yAmplitude*(\xPulse+\xEdge)/(\xPulse+2*\xEdge)});

\draw[aline,->] (\xC,\yStart) -- (\xC+\xAxisLen,\yStart);
\draw[aline,->] (\xC,\yStart) -- (\xC,\yStart+\yEnd);

\draw[line]
    (\xC+\xStart,\yStart)
    cos (\xC+\xStart+0.5*\xEdge,\yStart+0.5*\yAmplitude)
    sin (\xC+\xStart+\xEdge,\yStart+\yAmplitude)
    -- (\xC+\xStart+\xEdge+\xPulse,\yStart+\yAmplitude)
    cos (\xC+\xStart+\xEdge+\xPulse+0.5*\xEdge,\yStart+0.5*\yAmplitude)
    sin (\xC+\xStart+\xEdge+\xPulse+\xEdge,\yStart);

\node[nlabel] at (\xC+\xStart+\xEdge+0.5*\xPulse,\yStart+0.52*\yAmplitude) {$\displacement$};
\node[nlabel] at (\xC+\xStart+\xEdge+0.5*\xPulse,\yStart+\ylabel) {$t$};

\end{tikzpicture}